\documentclass[showpacs,preprintnumbers,amsmath,amssymb,nofootinbib]{revtex4}

\usepackage{graphicx,epsf, epsfig, psfig, amssymb}
\usepackage{bm} 

\def\nn{\nonumber}

\def\be{\begin{equation}}
\def\ee{\end{equation}}
\def\beq{\begin{eqnarray}}
\def\eeq{\end{eqnarray}}

\def\ii{{\rm i}}

\def\IL{\relax{\rm I\kern-.18em L}}

\def\nn{\nonumber}
\def\f{\frac}

\begin{document}

\title{Quasinormal modes and classical wave propagation in 
analogue black holes}

\author{Emanuele Berti}
\email{berti@iap.fr} \affiliation{McDonnell Center for the Space Sciences,
Department of Physics, Washington University, St. Louis, Missouri 63130, USA}

\author{Vitor Cardoso}
\email{vcardoso@teor.fis.uc.pt} \affiliation{Centro de F\'{\i}sica
Computacional, Universidade de Coimbra, P-3004-516 Coimbra,
Portugal}

\author{Jos\'e P. S. Lemos}
\email{lemos@fisica.ist.utl.pt} \affiliation{Centro
Multidisciplinar de Astrof\'{\i}sica - CENTRA, Departamento de
F\'{\i}sica, Instituto Superior T\'ecnico, Av. Rovisco Pais 1,
1049-001 Lisboa, Portugal}

\date{\today}

\begin{abstract}

Many properties of black holes can be studied using acoustic analogues
in the laboratory through the propagation of sound waves. We
investigate in detail sound wave propagation in a rotating acoustic
$(2+1)$-dimensional black hole, which corresponds to the ``draining
bathtub'' fluid flow.  We compute the quasinormal mode frequencies of
this system and discuss late-time power-law tails. Due to the presence
of an ergoregion, waves in a rotating acoustic black hole can be
superradiantly amplified. We compute superradiant reflection coefficients
and instability timescales for the acoustic black hole bomb, the
equivalent of the Press-Teukolsky black hole bomb. Finally we discuss
quasinormal modes and late-time tails in a non-rotating canonical
acoustic black hole, corresponding to an incompressible, spherically
symmetric $(3+1)$-dimensional fluid flow.

\end{abstract}


\maketitle

\section{Introduction}

Black holes are one of the most fascinating predictions of general
relativity. Being a solution of the classical Einstein field equations
in vacuum, they are the simplest objects which can be built out of
spacetime itself. Hawking \cite{hawking} showed that when quantum
effects are taken into account, black holes are not really black: they
slowly evaporate by emitting an almost thermal radiation. Hawking's
prediction has been tested time and again in very different ways. It
is now clear that the appearance of Hawking radiation does not depend
on the dynamics of the Einstein equations, but only on their
kinematical structure, and more specifically on the existence of an
apparent horizon \cite{visserlaws,visserhawking}. The discovery of
Hawking radiation uncovered a number of fundamental questions: among
them the information puzzle, the issue of the black hole final state,
and so on. For these reasons an experimental verification of the
Hawking effect would be of the utmost importance. Unfortunately
astrophysical black holes, having a temperature much smaller than the
temperature of the cosmic microwave background, accrete matter more
efficiently than they evaporate. An alternative could be provided by
the existence of large extra dimensions: if gravity is effective at
the TeV scale, black holes could be produced in particle accelerators
\cite{giddings}. However, even if these mini-black holes are actually
produced in colliders, it is unlikely that they will yield any
conclusive evidence of the existence of Hawking radiation.

Prospects for detecting Hawking radiation changed when Unruh
\cite{unruh} realized that the basic ingredients of Hawking radiation
can experimentally be reproduced in the laboratory. Since Hawking
radiation crucially depends on the existence of an apparent horizon,
the experimental setup should display the essential features that
define apparent horizons in general relativity. Unruh considered
precisely such a system: a fluid moving with a space-dependent
velocity, for example water flowing through a nozzle. Where the fluid
velocity exceeds the sound velocity we get the equivalent of an
apparent horizon for sound waves. This is the acoustic analogue of a
black hole, or a {\it dumb hole}. Following on Unruh's dumb hole
proposal many different kinds of analogue black holes have been
devised, based on condensed matter physics, slow light etcetera
\cite{novello,visser}. At present the Hawking temperatures associated
to these analogues are too low to be detectable, but the situation is
likely to change in the near future \cite{barcelo,unruh2}.

In order to detect Hawking radiation, a full understanding of the {\it
classical} physics of analogue black holes is necessary. Not only must
one control what happens in the experimental situation, but the
understanding of classical phenomena may bring clues on how to favor
the probabilities to detect Hawking radiation. It is also worth
stressing that some purely classical phenomena shed light on quantum
aspects of (analogue and general-relativistic) black hole physics. For
example, positive and negative norm mixing at the horizon leads to
non-trivial Bogoliubov coefficients in the calculations of Hawking
radiation \cite{corleyjacobson}; superradiant instabilities of the
Kerr metric are related to the quantum process of Schwinger pair
production \cite{schwinger}; and more speculatively (classical) highly
damped black hole oscillations could be related to area quantization
\cite{hod}.

In this paper we carry out a comprehensive study of wave propagation
in analogue black holes. We consider in detail two acoustic black hole
metrics: the ``draining bathtub'' model for a $(2+1)$-dimensional
fluid flow and the ``canonical'' metric for a non-rotating,
spherically symmetric $(3+1)$-dimensional acoustic black hole
\cite{visser}. For each acoustic metric we compute the characteristic
oscillation frequencies using a WKB approach. These complex
characteristic frequencies, called quasinormal modes (QNMs), play a
very important role in the classical physics of black holes. They
govern the late time behavior of waves propagating outside the black
hole: any perturbation of the black hole, after an initial transient,
will damp exponentially in the so-called ``ringdown phase''. The
frequencies and damping times of this ringdown signal depend only on
the black hole parameters, such as the mass, charge and angular
momentum \cite{kokkotas}, and can therefore be used to estimate black
hole parameters from observational data \cite{echeverria}. 
Therefore, in cases where a ``formal'' definition of (say) the mass
and angular momentum of acoustic black holes is missing, quasinormal
(QN) frequencies may be used to define these quantities {\it
operationally}. According to recent speculations, highly damped QNMs
may yield some information on quantum properties of a black hole. In
particular, it has been conjectured that highly damped QN frequencies
could be linked to black hole area quantization \cite{hod}. Motivated
by these conjectures we also study highly damped QNMs of acoustic
black holes. We find that for a $(2+1)$-dimensional acoustic black
hole there are no asymptotic QN frequencies, whereas for the
$(3+1)$-dimensional canonical black hole QN frequencies are
asymptotically given by $4\pi \omega =\log{[(3-\sqrt{5})/2]}
-\ii(2n+1)\pi$.
This result does not seem to support any of the recent
conjectures, but perhaps this is no surprise, since Hod's argument
relies heavily on black hole thermodynamics. Even the very formulation
of the laws of black hole thermodynamics for analogue black holes is a
non-trivial matter \cite{visserlaws}.

After the exponential decay characteristic of the ringdown phase,
black hole perturbations decay with a power-law tail \cite{price1} due
to backscattering off the background curvature. Here we compute the
late-time tails of wave propagation in the draining bathtub and
canonical acoustic black hole metrics. We show that for the
$(2+1)$-dimensional fluid flow the field falloff at very late times is
of the form $\Psi\sim t^{-(2m+1)}$, where $m$ is the angular quantum
number. This time exponent is characteristic of any
$(2+1)$-dimensional flow, and not just of a black hole. For the
$(3+1)$-dimensional canonical acoustic geometry, the field falloff is
of the form $\Psi \sim t^{-(2l+6)}$.

The rotating draining bathtub metric, possessing an ergoregion, can
display the phenomenon of superradiance
\cite{schutzhold,basak1,basak2}. We compute reflection coefficients
for this superradiant scattering by numerical integration of the
relevant equations. Enclosing the acoustic black hole by a reflecting
mirror we can exploit superradiance to destabilize the system, making
an initial perturbation grow exponentially with time: we have an
``acoustic black hole bomb'' \cite{pressteu,bhb,putten}. We compute
analytically and numerically the frequencies and growing timescales
for this instability.  An interesting feature of acoustic geometries
is that the acoustic black hole {\it spin} can be varied independently
of the black hole {\it mass}. Therefore, at variance with the Kerr
metric, the spin can be made (at least in principle) very large, and
rotational superradiance in acoustic black holes can be very
efficient. It should not be difficult to set up an experimental
apparatus to observe an acoustic black hole bomb in the lab.  A
concrete example of an experimental setup in which this idea can be
realized with moderate experimental effort has been fully worked out
by Sch\"utzhold and Unruh \cite{schutzhold}, and is based on the study
of gravity waves in shallow water.

The paper is organized as follows. In Section \ref{bathtub} we discuss
the $(2+1)$-dimensional draining bathtub metric. We first introduce
the formalism describing sound propagation in this acoustic metric,
and shortly describe the possible experimental setup (gravity waves in
a shallow basin). Then we compute QNMs, discuss late-time tails,
introduce the phenomenon of superradiant amplification and quantify
the timescales for the acoustic black hole bomb instability. In
Section \ref{canonical} we repeat the analysis for the canonical
$(3+1)$-dimensional acoustic black hole metric (of course, the absence
of an ergoregion means that we have no superradiance in this
case). The conclusions follow in Section \ref{conclusions}.

\section{Draining bathtub: a rotating acoustic black hole}\label{bathtub}

In this Section we consider a simple ``draining bathtub'' model, first
introduced in \cite{visser}, for a rotating acoustic black hole. We
write down the acoustic metric and the wave equation describing sound
propagation in this model, and describe a concrete example for a
possible experimental setup. We compute QNMs and discuss the late-time
tail behavior of the system. Due to the presence of an ergoregion,
sound waves in this acoustic black hole can be superradiantly
amplified. We quantify this amplification and discuss the possibility
to build an acoustic black hole bomb in the lab.

\subsection{Formalism and basic equations}

Consider a fluid having (background) density $\rho$.  Assume the fluid
to be locally irrotational (vorticity free), barotropic and
inviscid. From the equation of continuity, the radial component of the
fluid velocity satisfies $\rho v^r\sim 1/r$. Irrotationality implies
that the tangential component of the velocity satisfies $v^\theta\sim
1/r$. By conservation of angular momentum we have $\rho v^\theta\sim
1/r$, so that the background density of the fluid $\rho$ is
constant. In turn, this means that the background pressure $p$ and the
speed of sound $c$ are constants. The acoustic metric describing the
propagation of sound waves in this ``draining bathtub'' fluid flow is
\cite{visser}:
\begin{equation}
ds^2=
-\left (c^2-\frac{A^2+B^2}{r^2} \right )dt^2+\frac{2A}{r}drdt-2Bd\phi 
dt+dr^2+
r^2d\phi^2.
\label{metric1}
\end{equation}
Here $A$ and $B$ are arbitrary real positive constants related to the
radial and angular components of the background fluid velocity:
\be
{\vec v}=\frac{-A {\hat r}+B {\hat \theta}}{r}\,.
\ee
This flow velocity can be obtained as the gradient of a velocity
potential, $\vec v=\nabla \psi$, where
\be\label{vpot}
\psi=-A\log (r/a)+B\phi\,.
\ee
and $a$ is some (irrelevant) length scale.

In the non-rotating limit $B=0$ the metric (\ref{metric1}) reduces to
a standard Painlev\'e-Gullstrand-Lema\^itre type metric
\cite{PGL}. The acoustic event horizon is located at $r_H=A/c$, and
the ergosphere forms at $r_{ES}=(A^2+B^2)^{1/2}/c$.
Unruh \cite{unruh} first realized that the propagation of a sound wave
in a barotropic inviscid fluid with irrotational flow is described by
the Klein-Gordon equation $\nabla_{\mu}\nabla^{\mu}\Psi=0$ for a
massless field $\Psi$ in a Lorentzian acoustic geometry, which in our
case takes the form (\ref{metric1}). In our acoustic geometry we can
separate variables by the substitution
\be
\Psi(t,r,\phi)=R(r)e^{\ii (m\phi-\omega t)}\,.
\ee
Then we obtain a simple ordinary differential equation for the radial
variable:
\be
R_{,rr}+P_1(r)R_{,r}+Q_1(r)R=0\,,
\ee
where
\beq
P_1(r)&=&\f{A^2+r^2c^2+2\ii A(Bm-r^2\omega)}{r(r^2c^2-A^2)}\,,\nn\\
Q_1(r)&=&-\f{2\ii A Bm-B^2m^2+c^2m^2r^2+2Bm\omega r^2-r^4\omega^2}
{r^2(r^2c^2-A^2)}\,.
\eeq
Notice that if we take the incompressible fluid limit $c\rightarrow
\infty$ we get well known equations from fluid dynamics
\cite{chandra,drazin}. We now introduce a tortoise coordinate $r_*$
defined by the condition
\be
\frac{dr_*}{dr}=\Delta\,,
\ee
where $\Delta\equiv (1-A^2/c^2r^2)^{-1}$. Explicitly,
\be
r_*=r+\f{A}{2c}\log\left|\f{cr-A}{cr+A}\right|\,.
\ee
Setting $R=ZH$ we get:
\be\label{ZH}
Z\Delta^2 H_{,r_* r_*}
+\left[\Delta(2Z_{,r}+P_1Z)+\Delta'Z\right]H_{,r_*}+
\left(Z_{,rr}+P_1Z_{,r}+Q_1Z\right)H=0\,.
\ee
To obtain a Schr\"odinger-like equation we impose the coefficient of
$H_{,r_*}$ to be zero:
\be
Z_{,r}+\f{1}{2}\left[
\f{(c^2r^2-A^2)+2\ii A(Bm-r^2\omega)}{r(c^2r^2-A^2)}
\right]Z=0\,.
\ee
A solution of this equation can easily be found:
\be
Z=r^{1/2}\exp{\left[
\left(\f{\ii Bm}{A}-1\right)\ln r
+\f{\ii(A^2\omega-Bmc^2)}{2Ac^2}\ln (c^2r^2-A^2)
\right]}\,.
\ee
Replacing this solution in Eq. (\ref{ZH}) we finally get the wave equation:
\be
H_{,r_* r_*}+
\left\{\f{1}{c^2}\left(\omega-\f{Bm}{r^2}\right)^2-
\left(\f{c^2r^2-A^2}{c^2r^2}\right)
\left[\f{1}{r^2}\left(m^2-\f{1}{4}\right)+\f{5A^2}{4r^4 c^2}\right]\right\}
H=0\,.
\label{waveequation}
\ee

The derivation given here is very similar to the one in \cite{basak1},
but we have corrected some typos in that paper. In particular notice
that the potential is the same as equation (10) in \cite{basak1},
except for a factor $1/4$ in the last term.  Some physical properties
of our ``draining bathtub'' metric are more apparent if we cast the
metric in a Kerr-like form performing the following coordinate
transformation (where again we correct some typos in \cite{basak2}):
\begin{equation}
dt\rightarrow d\tilde{t}+\frac{Ar}{r^2c^2-A^2}dr\,,\qquad
d\phi\rightarrow d\tilde{\phi}+\frac{BA}{r(r^2c^2-A^2)}dr\,.
\label{coordtransf}
\end{equation} 
Then the effective metric takes the form
\begin{equation}
ds^2=
-\left(1-\frac{A^2+B^2}{c^2r^2} \right)c^2 d\tilde{t}^2+
\left(1-\frac{A^2}{c^2r^2} \right )^{-1}dr^2
-2B d\tilde{\phi}d\tilde{t}+r^2d\tilde{\phi}^2\,.
\label{metric2}
\end{equation}
Notice an important difference between this acoustic metric and the
Kerr metric: in the ($t,t$) component of the metric (\ref{metric2})
the parameters $A$ and $B$ appear as a {\it sum} of squares. This
means that, {\it at least in principle}, there is no upper bound for
the rotational parameter $B$ in the acoustic black hole metric,
contrary to what happens in the Kerr geometry. Separating variables by
the substitution
\be
\Psi(\tilde{t},r,\tilde{\phi})=\sqrt{r} H(r)e^{\ii (m\tilde{\phi}-\omega \tilde{t})}\,,
\ee
one can show that the radial function $H(r)$ is again a solution of
equation (\ref{waveequation}).

The wave equation (\ref{waveequation}) can be recast in a more
convenient form by the following rescaling: $\hat r=rA/c$,
$\hat{\omega}=\omega A/c^2$, $\hat{B}=B/A$. We get
\be
H_{,\hat r_* \hat r_*}+Q H=0\,,
\label{waveequation2}
\ee
where the generalized potential
\be\label{Qdef}
Q\equiv 
\left\{ \left(\hat{\omega}-\f{\hat{B}m}{\hat r^2}\right)^2-V \right\}\,,
\qquad
V\equiv \left(\f{\hat r^2-1}{\hat r^2}\right)
\left[\f{1}{\hat r^2}\left(m^2-\f{1}{4}\right)+\f{5}{4\hat r^4}\right]\,.
\ee

The rescaling effectively sets $A=c=1$ in the original wave equation,
and picks units such that the acoustic horizon $\hat r_H=1$. From now
on we shall omit hats in all quantities (unless otherwise stated). The
rescaled wave equation (\ref{waveequation2}) will be the starting
point of our analysis of QNMs, late-time tails and superradiant
phenomena. Before giving details of this analysis we describe a
possible experimental setup in which all of the classical physics we
are going to discuss could, at least in principle, be reproduced.

\subsection{A possible experimental setup}\label{gwanal}

The acoustic metric (\ref{metric1}) can be realized in a simple
experimental setup, that was described in detail by Sch\"utzhold and
Unruh \cite{schutzhold}. In this setup it no longer describes a sonic
analogue but rather a shallow basin gravity wave analogue of a black
hole.  The idea is to use gravity waves in a viscosity free,
incompressible liquid with irrotational flow: under appropriate
circumstances, one can envisage the use of common fluids like water or
mercury. Sch\"utzhold and Unruh assumed a shallow water, long
wavelength approximation: the gravity wave amplitude $\delta h$, their
wavelength $\lambda$ and the depth of the basin $h_B$ are such that
$\delta h\ll h_B\ll \lambda$. Relaxing the assumptions that the bottom
of the tank and the background flow surfaces are flat and parallel,
they showed that the most general rotationally symmetric and locally
irrotational background flow profile can be described precisely by the
draining bathtub metric (\ref{metric1}) when the radial slope of the
bottom of the tank -- that in cylindrical coordinates $(z,~r,~\phi)$
will be described by some function $f(r)$ -- is small: $f'(r)\ll
1$. In this gravity wave black hole analogue the constants $A$ and $B$
are proportional to the radial and tangential components of the
background flow velocity:
\be
v^\phi=\frac{B}{r^2}\,,
\qquad 
v^r=-\frac{A h_\infty}{rh\sqrt{1+f'(r)^2}}\,,
\ee
where $h_\infty$ is the height of the tank far from the black hole,
and the slope of the tank satisfies the relation
\be
f(r)=-\frac{(A^2+B^2)}{gr^2}\,.
\ee
In the previous equations $g$ is the gravitational acceleration,
related to the constant $c$ of the acoustic black hole metric
(\ref{metric1}) by the relation
\be
c=\sqrt{gh_\infty}\,.
\ee
One of the main advantages of this acoustic black hole is apparent
from this equation: the speed of the gravity waves can simply be tuned
to one's needs by adjusting the height of the basin
$h_\infty$. Another advantage, that we will not exploit here, is that
the inclusion of surface tension and viscosity can be used to
manipulate the waves' dispersion relation. To be concrete we will
sometimes consider the following plausible choice of physical
parameters, as suggested in \cite{schutzhold}: gravity waves of
amplitude $\delta h\sim 1$ mm and wavelength $\lambda\sim 10$ cm; a
tank of height $h_\infty\sim 1$ cm (so that $c\sim 0.31$ m~s$^{-1}$);
and a typical characteristic size of the acoustic black hole horizon
$r_H\sim 1$ m, corresponding to $A=c r_H\sim 0.31$ m$^2$~s$^{-1}$.

\subsection{Quasinormal modes}\label{dbqnms}

Numerical and analytical studies of a fairly general class of initial
data show that the evolution of the perturbations of a black hole
spacetime can roughly be divided in three parts: (i) The first part is
the prompt response at very early times. In this phase, which is the
obvious counterpart of light cone propagation, the form of the signal
depends strongly on the initial conditions. (ii) At intermediate times
the signal is dominated by an exponential decay, whose frequencies and
damping times are determined by the black hole QNMs. In this
``ringdown'' phase the signal depends entirely on the black hole
parameters (typically mass, charge and angular momentum). (iii) Due to
backscattering off the spacetime curvature, at late times the
propagating wave leaves a ``tail'' behind, usually a power law falloff
of the field. This power law seems to be highly independent of the
initial data, and persists even if there is no black hole horizon. In
fact it only depends on the asymptotic far region. In the following we
study the QNM ringing phase of our acoustic black hole metric; then we
will consider its late-time behavior.

The characteristic QNMs of the rotating acoustic black hole can be
defined in the usual way, imposing appropriate boundary conditions and
solving the corresponding eigenvalue problem. Close to the event
horizon the solutions of equation (\ref{waveequation2}) behave as
\begin{equation}
H \sim e^{\pm \ii\left (\omega-Bm\right)r_* }\,.
\label{bound1}
\end{equation}
Classically, only ingoing waves -- that is, waves falling into the
black hole -- should be present at the horizon. This means (according
to our conventions on the time dependence of the perturbations) that
we must choose the minus sign in the exponential. At spatial
infinity the solutions of (\ref{waveequation2}) behave as
\begin{equation}
H \sim e^{\pm \ii\omega r_*}\,.
\label{bound2}
\end{equation}
In this case we require that only outgoing waves (waves leaving the
domain under study) should be present, and correspondingly choose the
plus sign in the exponential. This boundary condition at infinity may
be cause for objections. Indeed, no actual physical apparatus will be
accurately described by these boundary conditions: a real acoustic
black hole experiment will certainly not extend out to
infinity. However, we may imagine using some absorbing device to
simulate the ``purely outgoing'' wave conditions at infinity (for
another example in which an absorbing device modeling spatial infinity
could be required, cf. Section XI of \cite{schutzhold} -- in
particular their Fig. 5). In any event, later on we shall consider the
alternative possibility of ``boxed'' boundary conditions, describing a
closed system. Boxed boundary conditions were also considered in
\cite{bhb}.

For assigned values of the rotational parameter $B$ and of the angular
index $m$ there is a discrete (and infinite) set of QN frequencies,
$\omega_{QN}$, satisfying the wave equation (\ref{waveequation2}) with
boundary conditions specified by Eqs.  (\ref{bound1}) and
(\ref{bound2}). The QN frequencies are in general complex numbers, the
imaginary part describing the decay or growth of the perturbation,
because the time dependence is given by $e^{-\ii \omega t}$. We expect
the black hole to be stable against small perturbations, and therefore
$\omega_{QN}$ is expected to have a negative imaginary part, so that
the perturbation decays exponentially as time goes by. We could not
prove stability of the draining bathtub metric in general, but we
managed to derive upper bounds on the frequencies of unstable modes
(if they exist at all). We give the derivation of these upper bounds
in Appendix \ref{unstable}. As usual, we will order the QN frequencies
$\omega_{QN}$ according to the absolute value of their imaginary part:
the fundamental mode (labeled by an integer $n=0$) will have the
smallest imaginary part (in modulus), and so on.

\subsubsection{Slowly decaying modes of non-rotating black holes}

The lowest QNMs control the ringing behavior of any classical
perturbation outside the black hole. Higher overtones, having a larger
imaginary part, are damped more quickly and play a negligible role. In
particular, it is known \cite{kokkotas} that the fundamental ($n=0$)
QNM effectively determines the response of the black hole to exterior
perturbations. For these slowly damped QNMs, a WKB approximation -- as
developed by Schutz, Will and others \cite{willwkb,will,kokkotaswkb}
-- is accurate enough. As a convergence check, we will sometimes
extend the WKB treatment to sixth order \cite{konoplyawkb}. The WKB
method demands that the generalized potential $Q$ defined by Eq.
(\ref{Qdef}) has a single maximum outside the horizon. For
non-rotating black holes ($B=0$) this is certainly true provided $m$
is not zero, and one can easily extract the lowest QN frequencies in
this case.  We show them in Table \ref{tab:fundnonrot}, where we also
show the convergence of the WKB scheme as the order of approximation
increases.  For $m=0$ the situation is not so simple, and we haven't
been able to extract the QN frequencies using this method, because the
potential possesses two extrema.  Due to the symmetry of the potential
(\ref{Qdef}), QN frequencies of non-rotating black holes are
independent of the sign of $m$ (to any positive QN frequency for
positive $m$ corresponds a negative QN frequency for negative $m$).

\begin{table}
\centering
\caption{\label{tab:fundnonrot} The fundamental ($n=0$) QN frequencies
for the non-rotating acoustic black hole, using three WKB
computational schemes.  $\omega _{QN}^{(1)}$ is the result for the QN
frequency using only the lowest approximation \cite{willwkb}, $\omega
_{QN}^{(3)}$ is the value obtained using 3rd order improvements
\cite{will}, and finally $\omega _{QN}^{(6)}$ was computed using 6th
order corrections \cite{konoplyawkb}. Notice how for $m>1 $ the three
schemes yield very similar answers.}
\vskip 12pt
\begin{tabular}{@{}c|c|c|c@{}}  
\hline
\hline
$m$ &$\omega _{QN}^{(1)}$  &$\omega _{QN}^{(3)}$   &$\omega _{QN}^{(6)}$\\
\hline
\hline
1  & 0.696-0.353i & 0.321-0.389i &0.427-0.330i\\
2  & 1.105-0.349i & 0.940-0.353i &0.945-0.344i\\
3  & 1.571-0.351i & 1.465-0.353i &1.468-0.352i \\
4  & 2.054-0.352i & 1.975-0.353i &1.976-0.353i \\
\hline 
\hline
\end{tabular}
\end{table}
From Table \ref{tab:fundnonrot} we see that the imaginary part is
nearly constant as a function of $m$, whereas the real part
approximately scales with $m$. Indeed, in the limit of large $m$ one
can show directly from the WKB formula \cite{willwkb} that
$\omega_{QN}$ behaves as
\begin{equation}
\omega \sim \frac{m}{2}-i\frac{2n+1}{2\sqrt{2}}\,,\,m \rightarrow \infty\,,
\label{largembehav}
\end{equation}  
and indeed already for $m=4$ this formula yields very good agreement
with the results shown in Table \ref{tab:fundnonrot}.  This very same
result can also be obtained using a P\"oschl-Teller fitting potential
\cite{ferrari}. Just to give an idea of the orders of magnitude
involved consider an $m=2$ mode: if we take the ``typical'' gravity
wave analogue experimental parameters of Section \ref{gwanal} we get
(for the fundamental QNM with $m=2$) a frequency $\omega_R=0.945
\times c^2/A=0.293$ Hz and a damping timescale $\tau=1/|\omega_I| =
1/0.344 \times A/c^2=9.37$ s.

\subsubsection{Slowly decaying modes of rotating black holes}

\begin{figure}
\centerline{\mbox{
\psfig{figure=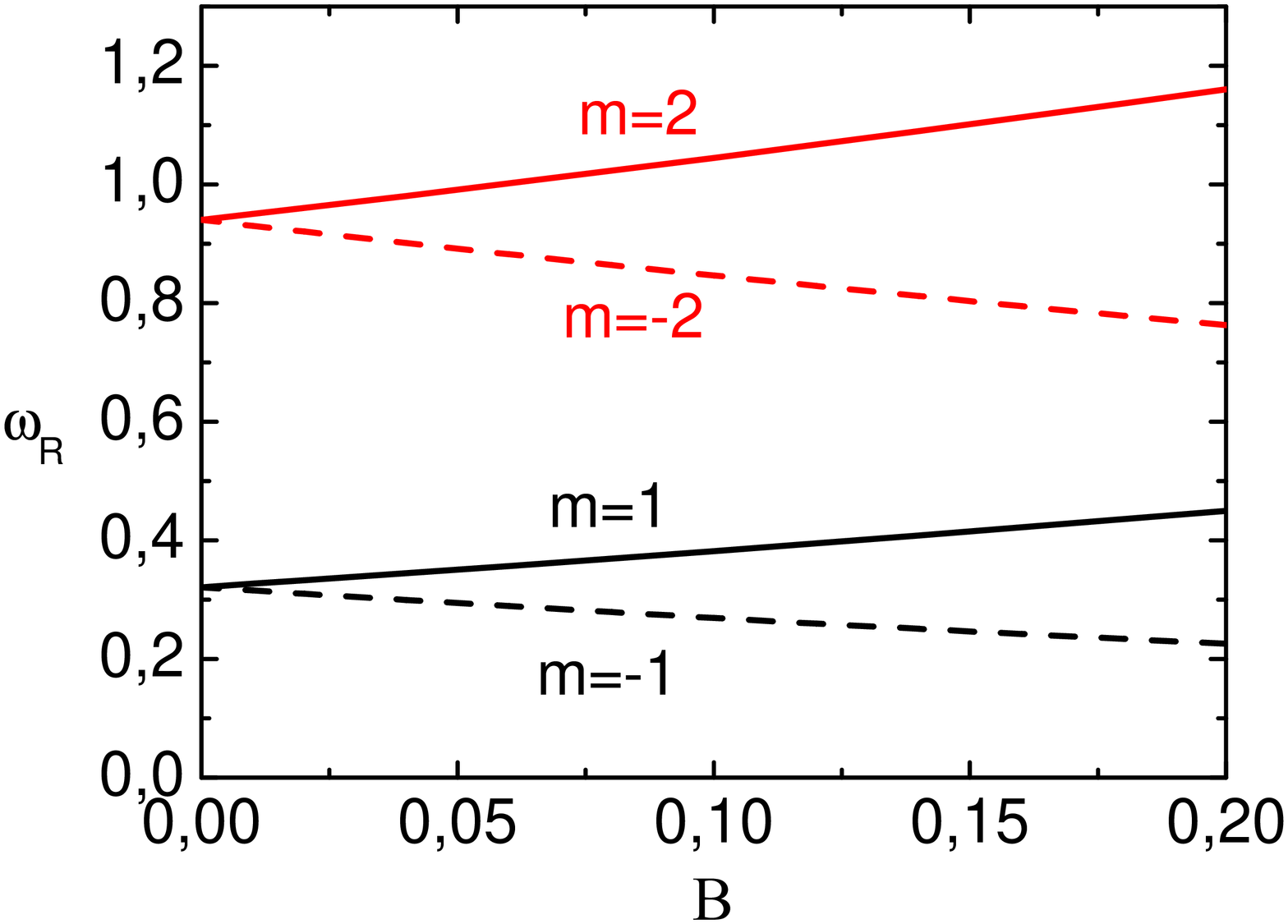,angle=270,width=9cm}
\psfig{figure=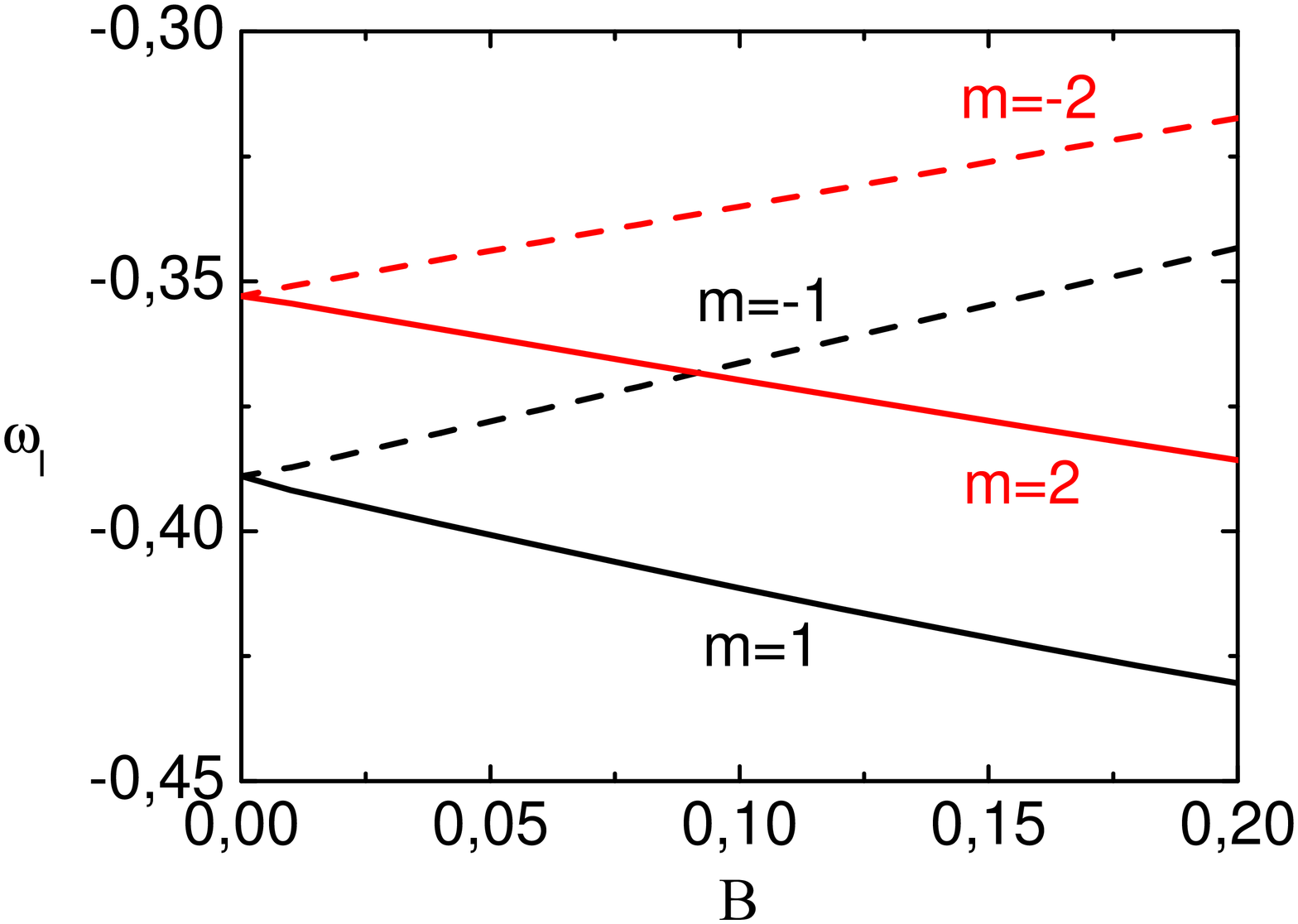,angle=270,width=9cm}
}}
\caption{ In the left panel we show the real part of the fundamental
QN frequency $\omega_R$ (and in the right panel we show the imaginary
part $\omega_I$) as a function of the rotation parameter $B$, for
selected values of $m$. For $m>0$ ($m<0$) $\omega_R$ and $|\omega_I|$
increase (decrease) with rotation. }
\label{fig:QNM.eps}
\end{figure}

For rotating black holes, the generalized potential has more than one
extremum. This immediately raises problems for the applicability of
the WKB technique.  Previous work on the Kerr geometry
\cite{will,kokkotaswkb} showed that the WKB method can still be used,
yielding good results, as long as the rotation parameter is small.
The way to handle the several extrema of the potential is the
following: start with $B=0$ and compute the roots of $dQ/dr_*=0$ (to
find the extrema; as we said before, for $B=0$ the generalized
potential $Q$ has a single maximum outside the horizon as long as
$m\neq 0$).  Compute $\omega _{QN}$ in the non-rotating case.  Then
add some small rotation $B$ to the black hole, and compute the roots
of $dQ/dr_*=0$. Now the roots will depend on $\omega$, but only one
root yields the correct non-rotating limit as $B\rightarrow 0$.  It is
this root that corresponds to the QNMs of the rotating black hole. Our
results, which can be trusted for $B\lesssim 0.2$, are shown in
Fig. \ref{fig:QNM.eps}. For $m>0$ ($m<0$) $\omega_R$ and $|\omega_I|$
increase (decrease) with rotation, at least in the range in which the
WKB method can be applied.  The QN frequency change with rotation is
not dramatic, but it can probably be used to apply a ``fingerprint
analysis'' of the acoustic black hole parameters {\it \'a la
Echeverria} \cite{echeverria}: that is, once we measure at least two
QNM frequencies we may infer the acoustic black hole parameters $A$ and $B$.
Carrying out such experiments in the lab can shed some light on the
applicability of similar ideas to test the no-hair theorem in the
astrophysical context \cite{dreyerfinn}.  A more complete analysis of
the QNMs is still needed to probe the high rotation regime: the
results shown in Fig. \ref{fig:QNM.eps} seem to indicate that
instability may set in for large $Bm$.  A continued fraction analysis
could be used to test this hypothesis, but it is beyond the scopes of
this paper \cite{cardoso4}.
\vskip 1mm

\subsubsection{Highly damped modes}

QNMs with a large imaginary part, i.e. with a very large overtone
number $n$, have recently become a subject of intense scrutiny for
black holes in general relativity and similar theories. The interest
in these modes comes from Hod's proposal \cite{hod} that they could be
related to black hole area quantization, and from Dreyer's suggestion
that a similar argument could be used to fix the Barbero-Immirzi
parameter in Loop Quantum Gravity \cite{dreyer}.  In this context, an
analytical calculation of highly damped QNMs was first carried out by
Motl for the Schwarzschild black hole \cite{motl1}. Subsequently Motl
and Neitzke \cite{motl2} used a complex-integration technique to
compute highly-damped QNMs of the Schwarzschild and
Reissner-Nordstr\"om black holes. Their analytical results are in
striking agreement with numerical data \cite{num1,num3} and
alternative analytical calculations \cite{nils}. The complex
integration method has also been generalized with success to other
black hole geometries \cite{ricardo,tamaki}, 
yielding again predictions in
excellent agreement with the numerical results \cite{num2}. In view of
this, we shall now build on the results and techniques of \cite{motl2}
to compute highly damped QNMs of a rotating acoustic black hole.

Let us first consider a non-rotating black hole. In the limit $B\to
0$ the wave equation (\ref{waveequation2}) reduces to
\be
\frac{d^2H}{dr_*^2}+
\left [ \omega ^2-
V(r)
\right] H=0\,,
\ee
where the potential
\be
V(r)\equiv \left ( 1-\frac{1}{r^2}\right ) 
\left (\frac{m^2-1/4}{r^2}+\frac{5}{4r^4} \right )\,.
\ee
Following (\cite{motl2}), we can determine the highly damped QNMs
looking at the behaviour of the potential near the singular point
$r=0$. In our case the leading term as $r\to 0$ is
\be
V \sim -\frac{5}{4r^6}\,,
\ee
and in the same limit we have $r_* \sim -r^3/3$. Thus 
\be
V \sim -\frac{5}{36r_*^2}=\frac{j^2-1}{4r_*^2}\,,
\ee
for $j=2/3$ (this is precisely the same power-law behavior found in
the case of the Schwarzschild black hole, except for the different
value of $j$). The result in \cite{motl2} carries over directly:
\be
e^{4\pi \omega}=-(1+2\cos{\pi j})\,.
\ee
However now $(1+2\cos{\pi j})=0$. This means that there are no
asymptotic QN frequencies for this black hole. A similar situation
occurs for electromagnetic perturbations of a five-dimensional black
hole \cite{num2}. If we add rotation to the black hole, i.e., if $B$
is non-zero, a similar analysis \cite{neitzke} implies that the
asymptotic QN frequencies behave as
\be
e^{4\pi (\omega-mB)}=-(1+2\cos{\pi j})\,,
\ee
where again $j=2/3$. Thus also in this case there will be no
asymptotic QN frequencies. 

This result is quite puzzling. It could mean either that the real part
grows without bound, or that it just doesn't converge to any finite
value. In any case, an application of Hod's conjecture to these black
holes seems impossible: the very definition of the classical laws of
black hole thermodynamics is non-trivial in the analogue case. In
fact, as argued by Visser \cite{visserlaws}, the laws of black hole
thermodynamics arise solely from the Einstein equations for the
metric; from a different perspective, it has been shown that the
Einstein equations themselves can be derived from purely
thermodynamical arguments \cite{TedEinsteinEOS}. A well-known
``weakness'' of analogue models is the fact that they can be used to
reproduce the kinematical aspects of the Einstein equations, but not
their dynamical aspects (see eg. \cite{BLSV} for the implications of
this distinction on the causal structure of acoustic
spacetimes). Hod's arguments \cite{hod} assume a thermodynamic
relation between black hole surface area and entropy: even if Hod's
conjecture is true, the absence of any such relation for analogues
could explain the missing link between the QNM spectrum and area
quantization.

\subsection{Late-time tails}

After the exponential QNM decay characteristic of the ringdown phase
black hole perturbations usually decay with a power-law tail
\cite{price1}, due to the backscattering of waves off the background
curvature.  The existence of late-time tails in black hole spacetimes
is by now well established. A strong body of evidence comes from
analytical and numerical calculations using linear perturbation theory
and even non-linear evolutions, for massless or massive fields
\cite{price1,price2,ching1,ching2,cardosoDdimtails}. In a very
complete analysis, Ching, Leung, Suen and Young \cite{ching1,ching2}
considered the late-time tails appearing when one deals with evolution
equations of the form (\ref{waveequation2}), and the potential $V$ is
of the form
\begin{equation}
V(r_*) \sim \frac{\nu(\nu+1)}{r_*^2}+
\frac{c_1\log{r_*}+c_2}{r_*^{\alpha}}\,,\qquad r_*\rightarrow \infty.
\label{potching}
\end{equation}
By a careful study of the branch cut contribution to the associated
Green's function they concluded that in general the late-time behavior
is dictated by a power-law or by a power-law times a logarithm. The
exponents of the power-law depend on the leading term at very large
spatial distances.  The case of interest for us here is when
$c_1=0$. Their conclusions, which we will therefore restrict to the
$c_1=0$ case, are (see Table 1 in \cite{ching1} or \cite{ching2}):
\newline 
(i) if $\nu$ is an integer the term $\nu(\nu+1)/r_*^2$ does not
contribute to the late-time tail. Since this term represents just the
pure centrifugal barrier, characteristic of flat space, one can expect
that indeed it does not contribute, at least in four-dimensional
spacetimes. Therefore, for integer $\nu$, it is the $c_2/r_*^{\alpha}$
term that contributes to the late-time tail. In this case the authors
of \cite{ching1,ching2} find that the tail is given by a power-law,
\begin{equation}
\Psi \sim t^{-\mu}\,\,,\,\,\mu>2\nu+\alpha\,\,,\,\, \alpha\, 
{\rm odd}\,\,{\rm integer}<2\nu+3.
\label{tailintnu1}
\end{equation}
where the exponent $\mu=2\nu+2\alpha-2$.  For all other real $\alpha$,
the tail is
\begin{equation}
\Psi \sim t^{-(2\nu+\alpha)}\,\,,\,\,{\rm all}\,\,{\rm other}\,\, 
{\rm real}\,\,\alpha.
\label{tailintnu2}
\end{equation}
\newline 
(ii) if $\nu$ is not an integer, then the main contribution to the
late-time tail comes from the $\nu(\nu+1)/r_*^2$ term. In this case
the tail is
\begin{equation}
\Psi \sim t^{-(2\nu+2)}\,\,,\,\,{\rm non\,integer}\,\,\nu.
\label{tailnonintnu}
\end{equation}
As an example of non-integer $\nu$, Cardoso {\it et al.}
\cite{cardosoDdimtails} have shown that the late-time tails of wave
propagation appearing in odd dimensional spacetimes do not depend on
the presence of the black hole at all.  For odd dimensions the
power-law is determined not by the presence of the black hole, but by
the very fact that the spacetime is odd dimensional. In this case the
field decays as
\be
\Psi \sim t^{-(2l+D-2)}\,,
\label{tailsD}
\ee
where $l$ is the angular index determining the angular dependence of
the field, and $D$ the number of spacetime dimensions. One can show
directly from the flat space Green's function that such a power-law is
indeed expected in flat, odd dimensional spacetimes.

From the aforementioned arguments we should expect late-time tails in
our draining bathtub acoustic black hole to be directly related to the
dimensionality of the underlying flow, and not to the presence of a
black hole, since we are dealing with a $(2+1)$-dimensional
flow. Indeed, at leading order the potential behaves as
\be
V \sim \frac{m^2-1/4}{r_*^2}=\frac{\nu(\nu+1)}{r_*^2}\,,
\label{as1}
\ee
where $\nu=m-1/2$.  We are thus in case (ii) above, because $\nu$ is
not an integer. So any perturbation in the vicinities of this black
hole will die out as a late-time tail of the form
\be
H \sim t^{-(2m+1)}\,.
\label{latet}
\ee
This power-law is just the one given by the general expression
(\ref{tailsD}), if one substitutes $D=3$.

Strictly speaking, the asymptotic form (\ref{as1}) for the potential
is valid only for non-rotating acoustic black holes. For the general
rotating case the leading term in the potential is $\omega$-dependent:
\be
V \sim \frac{m^2-1/4}{r_*^2}=\frac{\nu(\nu+1)+2\omega Bm}{r_*^2}\,.
\label{as2}
\ee
Since late-times are associated with small frequencies, {\it and} the
term $\nu(\nu+1)$ gives a contribution to the late-time tail, then it
follows that the late-time tails of rotating acoustic black holes are
the same as those for non-rotating black holes, expression
(\ref{latet}).

\subsection{Superradiance}\label{integration}

Superradiance is a general phenomenon in physics. Inertial motion
superradiance has long been known \cite{ginzburg}, and refers to the
possibility that a (possibly electrically neutral) object endowed with
internal structure, moving uniformly through a medium, may emit
photons even when it starts off in its ground state. Some examples of
inertial motion superradiance include the Cherenkov effect, the Landau
criterion for disappearance of superfluidity, and Mach shocks for
solid objects travelling through a fluid (cf. \cite{bekschiffer} for a
discussion). Non-inertial rotational motion also produces
superradiance. This was discovered by Zel'dovich \cite{zeldovich}, who
pointed out that a cylinder made of absorbing material and rotating
around its axis with frequency $\Omega$ can amplify modes of scalar or
electromagnetic radiation of frequency $\omega$, provided the
condition
\be\label{suprad}
\omega<m\Omega
\ee
(where $m$ is the azimuthal quantum number with respect to the axis of
rotation) is satisfied. Zel'dovich realized that, accounting for
quantum effects, the rotating object should emit spontaneously in this
superradiant regime. He then suggested that a Kerr black hole whose
angular velocity at the horizon is $\Omega$ will show both
amplification and spontaneous emission when the condition
(\ref{suprad}) for superradiance is satisfied. This suggestion was put
on firmer ground by a substantial body of work \cite{superr}. In
particular, it became clear that (even at the purely {\it classical}
level) superradiance is required to satisfy Hawking's area theorem
\cite{beksuperr,pressteu}.

Superradiance is essentially related to the presence of an ergosphere,
allowing the extraction of rotational energy from a black hole through
a wave equivalent of the Penrose process \cite{penrose}. Under certain
conditions, superradiance can be used to induce instabilities in Kerr
black holes \cite{schwinger}.  Indeed, all spacetimes admitting an
ergosphere and {\it no horizon} are unstable due to rotational
superradiance. This was shown rigorously in \cite{friedman}, but the
growth rate of the instability is too slow to observe it in an
astrophysical context \cite{cominsschutz}. Kerr black holes are
stable, but if enclosed by a reflecting mirror they can become
unstable due to superradiance \cite{ptbhb,bhb}; we will discuss this
``black hole bomb'' instability as applied to the analogue acoustic
black hole.

The possibility to observe rotational superradiance in analogue black
holes was considered by Sch\"utzhold and Unruh \cite{schutzhold}, and
more extensively by Basak and Majumdar \cite{basak1,basak2}, who
computed analytically the reflection coefficients in the low frequency
limit $\omega A/c^2\ll 1$. In particular, the authors of
\cite{schutzhold} showed that the ergoregion instability in gravity
wave analogues is related to the existence of an ``energy function''
[their Eq. (68)] that is not positive definite inside the
ergosphere. In the context of analogues, inertial superradiance based
on superfluid $^3$He has been studied by Jacobson and Volovik
\cite{tedvolovik}.

Here we present a quantitative calculation of the efficiency of
superradiant amplification for the draining bathtub metric. We work in
the frequency representation, and consider an incident plane wave of
unit amplitude at infinity, frequency $\omega$ and azimuthal index
$m$. Part of this wave will be reflected back by the medium, the
reflection coefficient being some (complex) number $R_{\omega m}$. In
terms of the wave equation (\ref{waveequation2}), this determines the
following boundary condition at infinity:
\be 
H\sim R_{\omega m} e^{\ii \omega r_*}+e^{-\ii \omega r_*}
\,\,,r\rightarrow \infty\,.
\label{asymptbeh}
\ee
At the sonic horizon ($r\to 1$, $r_*\to -\infty$) the solution behaves
like
\be 
H\sim T_{\omega m} e^{-\ii (\omega-mB) r_*}\,\,,r\rightarrow 1\,.
\label{asymptbeh1}
\ee
where $T_{\omega m}$ is the transmission coefficient.
An easy way to prove the existence of superresonance is to compute the
Wronskian of a solution of equation (\ref{waveequation2}) and of its
adjoint at the sonic horizon and at infinity.  From the constancy of
the Wronskian as a function of the radial coordinate, using the
boundary conditions (\ref{asymptbeh}) and (\ref{asymptbeh1}) we find
the following ``energy conservation'' condition:
\be
1-\left|R_{\omega m}\right|^2=
\left(1-\f{mB}{\omega}\right)\left|T_{\omega m}\right|^2\,.
\ee
Therefore, if $\omega <mB$ the reflection coefficient $\left|R_{\omega
m}\right|^2>1$, i.e. we have superradiance.

To compute numerically with improved accuracy the reflection
coefficient $R_{\omega m}$, we have used a refined condition at the
horizon:
\be\label{horizon}
H\sim T_{\omega m} e^{-\ii (\omega-mB) r_*}\left[1+y_1(r-1)+\dots\right]\,, 
\ee 
where the leading-order coefficient is
\be
y_1=\frac{(1+2B^2)m^2-2\omega Bm+1}{ 2[\ii(mB-\omega )+1]}\,.
\ee
We also keep higher-order terms to extract the reflection coefficient
at infinity. Namely, for the outgoing wave we use an expansion of the
form:
\be\label{outg}
H\sim e^{\ii \omega r_*}\left[1+\frac{z_1}{r}+\frac{z_2}{r^2}+\dots\right]
\,, 
\ee
where
\be
z_1=\ii\frac{B+(4m^2-1)}{ 8\omega}\,,\qquad
z_2=\ii z_1\left[\frac{Bm}{2}+\frac{(4m^2-9)}{16\omega}\right]\,,\nn
\ee
and for the ingoing wave we use the complex conjugate of
Eq. (\ref{outg}).  As a check of our code we have reproduced results
by Andersson {\it et al.} \cite{alp} for the superradiant
amplification of a scalar field in the vicinities of a Kerr black
hole. Our results are in perfect agreement with Fig. 1 of
\cite{alp}. In particular, we find a maximum amplification coefficient
given by $1-|R_{\omega m}|^2\simeq 0.2$ \% for $l=m=2$ scalar
perturbations of a near-extremal extremal Kerr black hole (in early
numerical work \cite{ptbhb,pressteu} the maximum amplification was
found to be $1-|R_{\omega m}|^2\simeq 0.3$ \%).

\begin{figure}
\centerline{\mbox{
\psfig{figure=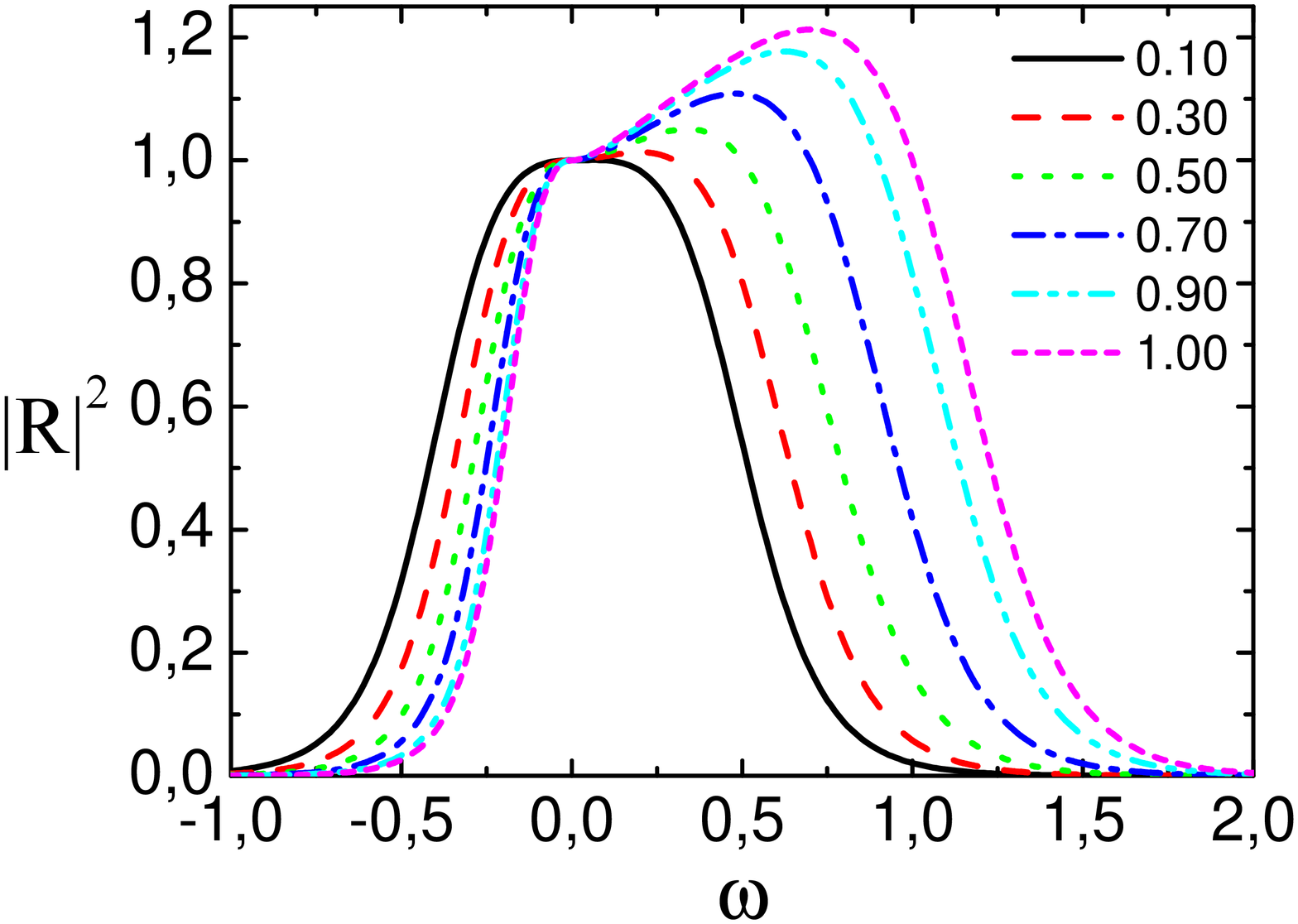,angle=270,width=9cm}
\psfig{figure=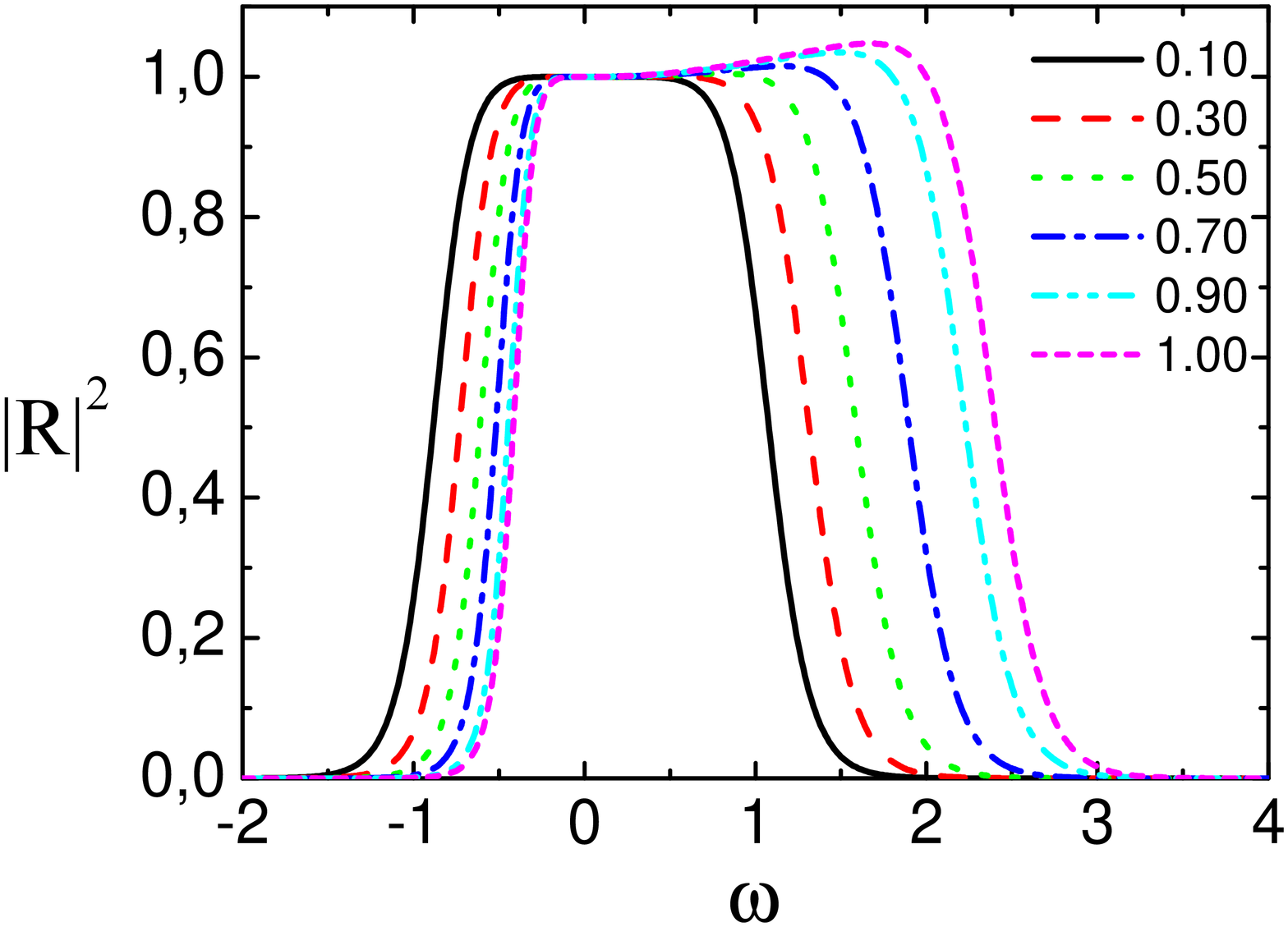,angle=270,width=9cm}
}}
\centerline{\mbox{
\psfig{figure=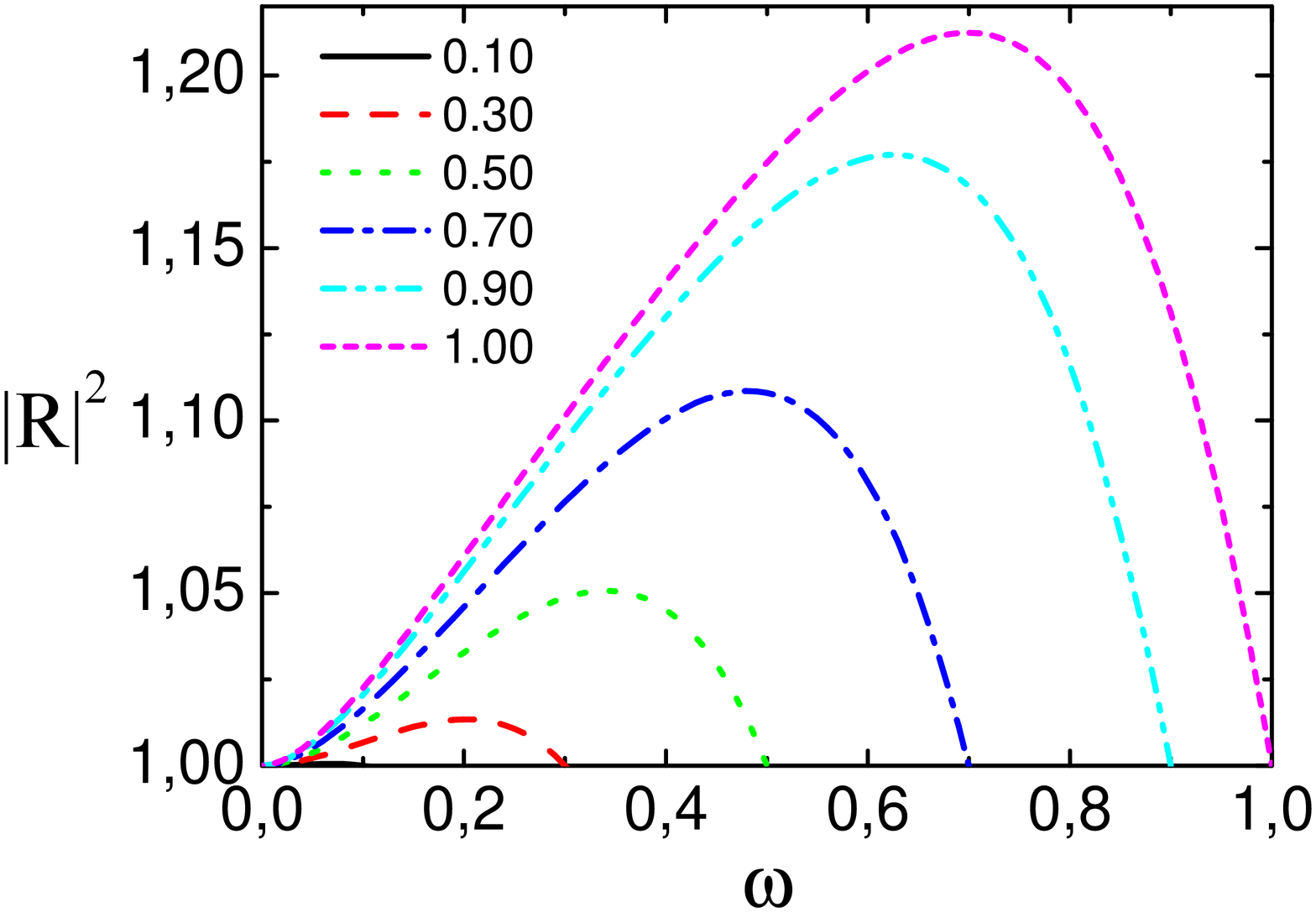,angle=270,width=9cm}
\psfig{figure=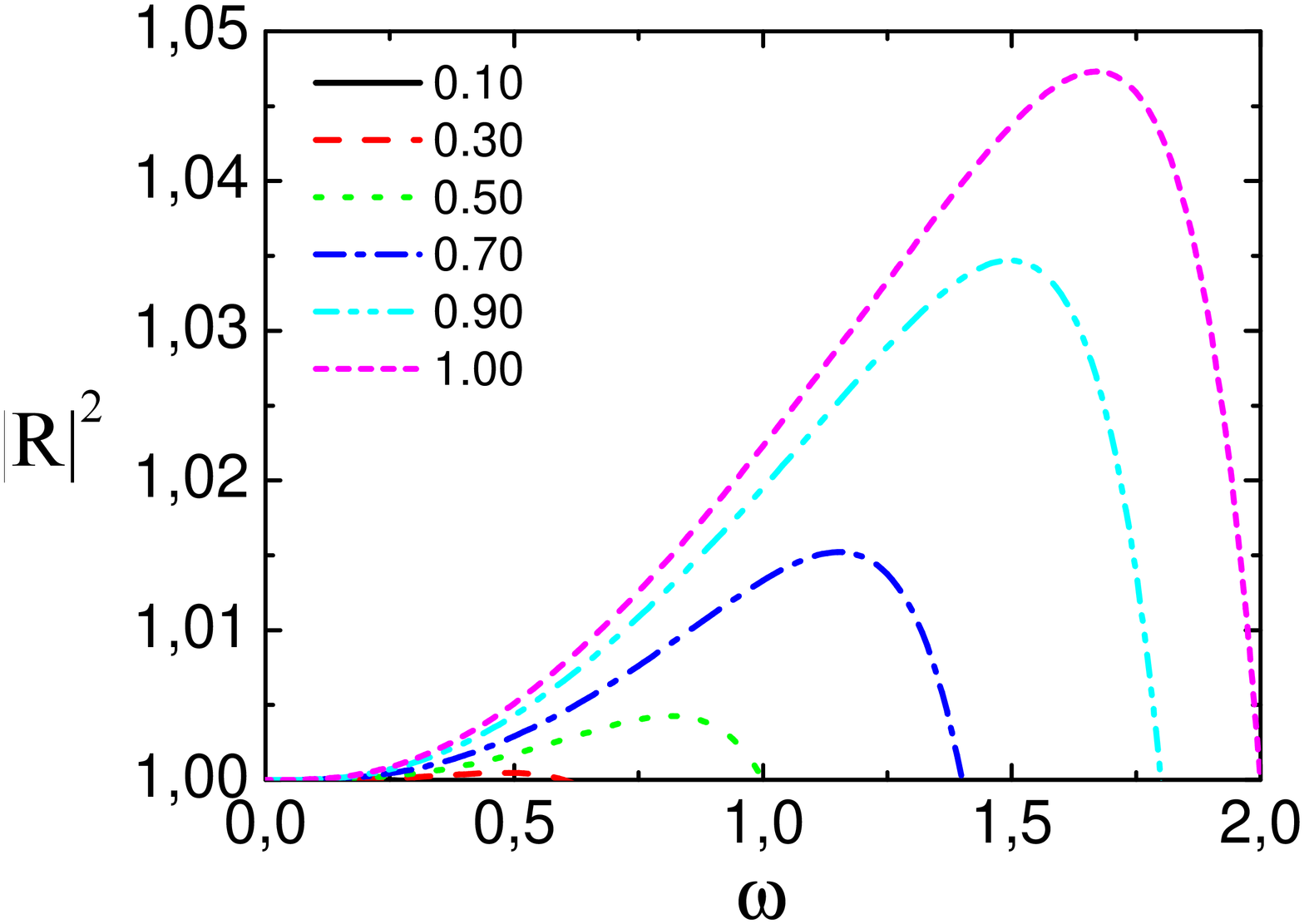,angle=270,width=9cm}
}}
\centerline{\mbox{
\psfig{figure=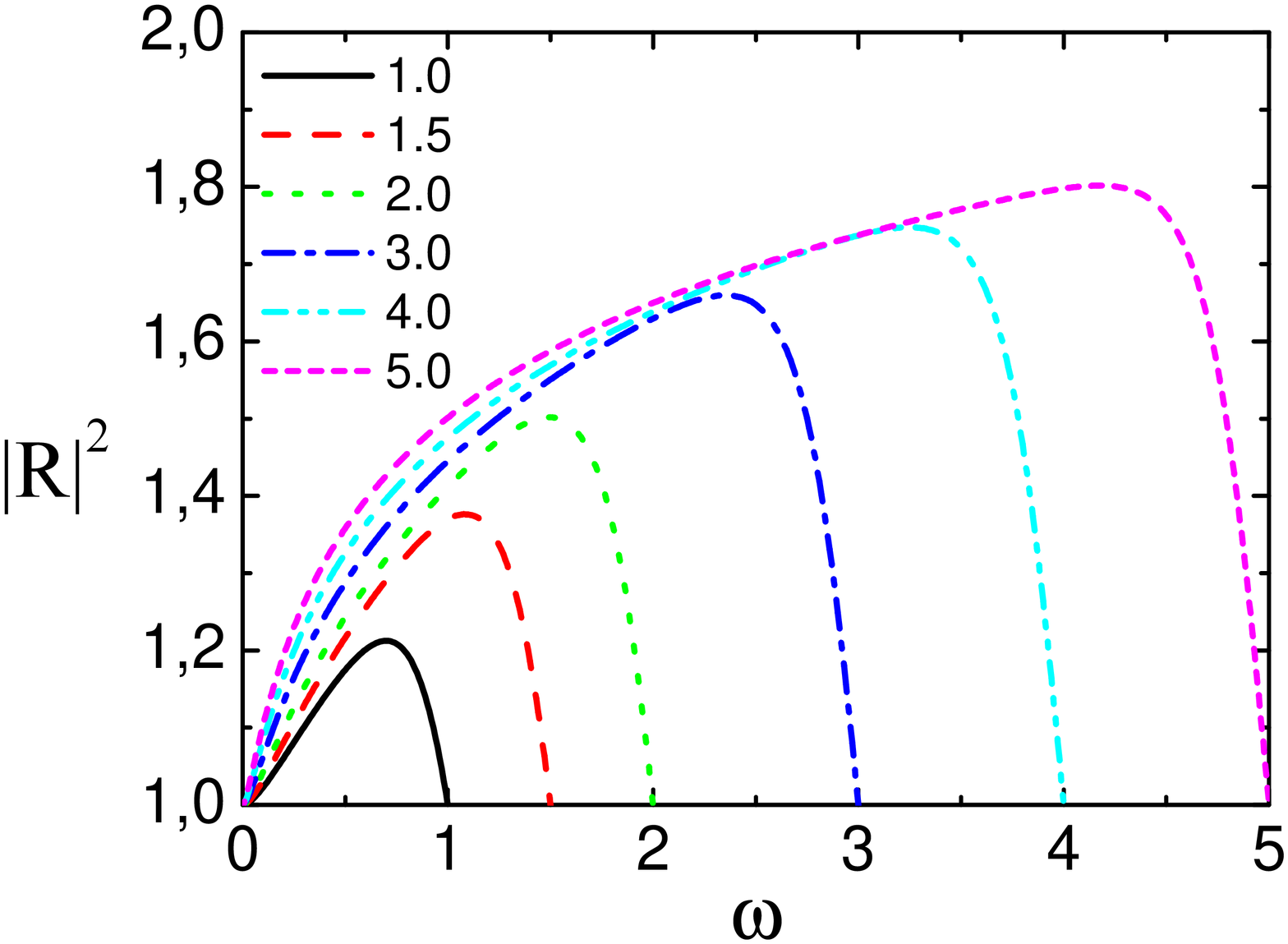,angle=270,width=9cm}
\psfig{figure=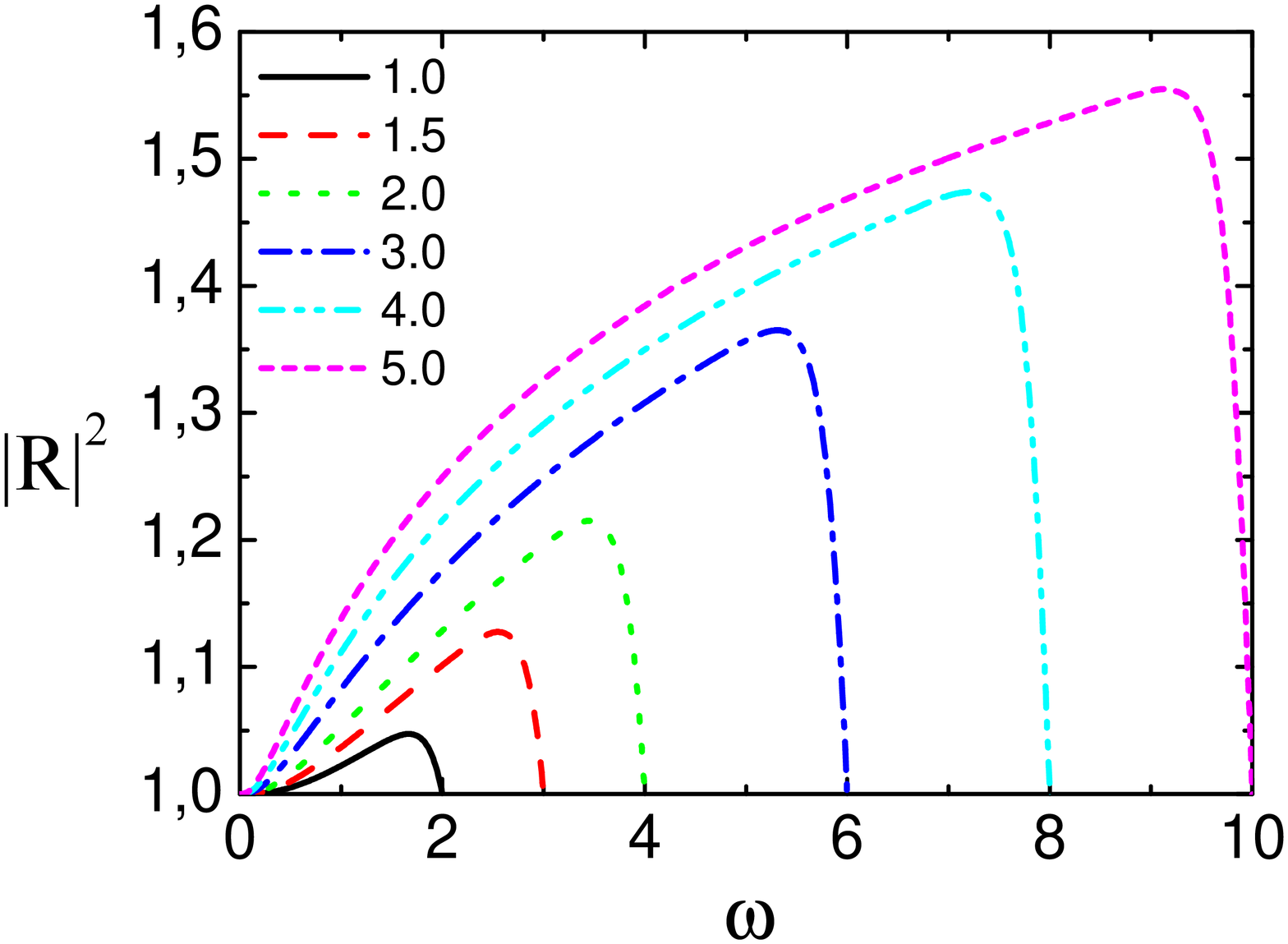,angle=270,width=9cm}
}}
\caption{ Reflection coefficient $\left|R_{\omega m}\right|^2$ as a
function of $\omega$ for $m=1$ (left panels) and $m=2$ (right
panels). Each curve corresponds to a different value of $B$, as
indicated. The top panels show that the reflection coefficient decays
exponentially at the critical frequency for superradiance,
$\omega_{SR}=mB$. The middle panels show a close-up view in the
superradiant regime for $B<1$: at $B=1$ the maximum amplification is
21.2 \% ($m=1$) and 4.7 \% ($m=2$). The bottom panels show that
superradiant amplification can become much more efficient for values
of the rotation parameter $B>1$. }
\label{fig:superradiantfactor}
\end{figure}

Results of the numerical integrations for the draining bathtub metric
are shown in Fig. \ref{fig:superradiantfactor}. Panels on the left
show the reflection coefficient $|R_{\omega 1}|^2$ for $m=1$, and
panels on the right show $|R_{\omega 2}|^2$ for $m=2$, for selected
values of the black hole rotation $B$. Panels on top show that, as
expected, in the superradiant regime $0<\omega<mB$ the reflection
coefficient $|R_{\omega m}|^2\geq 1$. Furthermore, as one increases
$B$ the reflection coefficient increases, and for fixed $B$, the
reflection coefficient $\left |R_{\omega m}\right|^2$ attains a
maximum at $\omega \sim mB$, after which it decays exponentially as a
function of $\omega$ outside the superradiant interval. This is very
similar to what happens when one deals with massless fields in the
vicinities of rotating Kerr black holes \cite{pressteu}. In particular,
from the close-up view in the middle panels we see that, for $B=1$,
the maximum amplification is 21.2 \% ($m=1$) and 4.7 \% ($m=2$).

As a final remark, and as we have anticipated, an important difference
between the acoustic black hole metric and the Kerr metric is that in
the present case there is no mathematical upper limit on the black
hole's rotational velocity $B$. In the bottom panels we show that,
considering values of $B>1$, we can indeed have larger amplification
factors for acoustic black holes.

Summarizing: if we are clever enough to build in the lab an acoustic
black hole that spins very rapidly, rotational superradiance can be
particularly efficient in analogues. This is an important result,
considering that the detection of rotational superradiance in the lab
is by no means an easy task, as originally predicted by Zel'dovich
\cite{zeldovich} and confirmed by recent reconsiderations of the
problem \cite{bekschiffer}. Of course, in any real-world experiment
the maximum rotational parameter will be limited. At the mathematical
level, the equations describing sound propagation (which are written
assuming the hydrodynamic approximation) will eventually break
down. Physically, if the angular component of the velocity $v^\theta$
becomes very large the dispersion relation for the fluid will change,
invalidating the assumptions under which we have derived our acoustic
metric \cite{schutzhold}. To be more concrete, let us consider again
the gravity wave analogue described in Section \ref{gwanal}. Then we
can use a very simple argument to limit the acoustic black hole
spin. The derivation of the rotating acoustic black hole metric is
based on the assumption that the bottom of the tank should not be too
steep, that is, $f'(r)\ll 1$. This condition translates into a
condition for $B$:
\be\label{Blimit}
B\ll\left(\frac{gr^3}{2}-A^2\right)^{1/2}\,,
\ee
which must be satisfied for all values of $r$, and in particular at
the acoustic black hole horizon $r=r_H$ (notice that here, and here
only, we have switched back to physical units). Setting $r=r_H$ in the
previous inequality and using the physical parameters quoted in
Section \ref{gwanal} we get the very stringent condition that $B\ll
2.2$ m$^2$~s$^{-1}$, or (in the dimensionless units we use throughout
this paper) $\hat B\ll 7.0\,$. In other words, we can only get values
of the rotation parameter larger than $\hat B\sim 1$ when the slope of
the tank is so large that the assumptions underlying the derivation of
the acoustic metric are not valid any more. Of course, this example
does not mean that such a constraint applies to {\it every} possible
experimental realization of the draining bathtub metric. However, it
serves as an illustration of the kind of experimental difficulties we
may expect to encounter in practice.

The superradiant phenomena we have described are purely {\it
classical} in nature. However, an interesting suggestion to observe
{\it quantum} effects in acoustic superradiance was put forward in
\cite{basak2}. To write down our acoustic metric we required the flow
to be irrotational and nonviscous. As a natural choice, we could use a
fluid which is well known to possess precisely these properties:
superfluid HeII. In this case the presence of vortices with quantized
angular momenta may lead to a quantized energy flux. The heuristic
argument presented in \cite{basak2} goes as follows. Let us imagine
that our black hole is a vortex with a sink at the centre. In the
quantum theory of HeII the wavefunction is of the form
$\Psi=\exp\left[\ii \sum_j \phi(\vec r_j) \Phi_{\rm ground}\right]$,
where $\vec r_j$ is the position of the $j$-th particle of HeII. The
velocity at any point is given by the gradient of the phase at that
point, $\vec v=\nabla \phi$, so that (roughly speaking) the velocity
potential (\ref{vpot}) can be identified with the phase of the
wavefunction. This phase will be singular at the sink
$r=0$. Continuity of the phase around a circle surrounding the sink
requires that the change of the wavefunction satisfies $\Delta
\phi=2\pi B$. For the wavefunction to be single valued, $B$ (that is,
the black hole's angular velocity at the horizon) must be the integer
multiple of some minimum value $\Delta B$, i.e., $B=n\,\Delta B$. Then
the angular momentum of the acoustic black hole would be forced to
change in integer multiples of $\Delta B$. Correspondingly, the
spectrum of the reflection coefficients may be given by equally-spaced
peaks with different strengths. This discrete amplification could
enhance chances of observing superradiance in acoustic black holes,
and rule out (or provide empirical support to) some of the many
competing heuristic approaches to black hole quantization.

\subsection{Superradiant instabilities: the acoustic black hole bomb}

The very existence of an ergoregion in the acoustic black hole metric
allows immediately for the possibility to make the system
unstable. Suppose we enclose the system by a reflecting mirror at
constant radius $r_0$.  Now, throw in a wave having frequency $\omega=
\omega_R+{\rm i}\, \omega_I$ such that $\omega_R<m\Omega$: the wave
will be amplified at the expense of the black hole's rotational energy
and travel back to the mirror. There it will be reflected and move
again towards the black hole, this time with increased
amplitude. Through repeated reflections, the waves' amplitude will
grow exponentially with time. This ``black hole bomb'' was first
proposed by Press and Teukolsky \cite{ptbhb}, and recently it has been
studied in detail by some of us \cite{bhb} in the context of the Kerr
metric. The detailed analysis performed in \cite{bhb} showed that the
black hole bomb can be characterized by a set of complex resonant
frequencies, the Boxed QuasiNormal Modes (BQNMs). The real part of a
BQNM is, not surprisingly, proportional to $1/r_0$, where $r_0$ is the
mirror radius: this is essentially the condition for the existence of
standing waves in the region enclosed by the mirror. The imaginary
part of the BQNMs is proportional to $(\omega_R-m\Omega)$: the system
can only become unstable if it is in the superradiant regime. Combined
with the standing wave condition this implies that the mirror should
be placed at some radius $r_0\gtrsim 1/m\Omega$ in order for the
system to be unstable. As rotational energy is extracted from the
system the black hole will spin down. For any given $r_0$ the
instability will eventually shut off, as the condition $r_0\gtrsim
1/m\Omega$ will no longer be satisfied.

An additional reason to study the system ``acoustic black
hole+mirror'' is related to the problem of boundary conditions at
infinity for QNMs (cf. Section \ref{dbqnms}). Whatever analogue model
we consider, no physical apparatus will ever extend out to
infinity. We are left with two possibilities: we can either use some
absorbing device to simulate spatial infinity, or we can impose
alternative boundary conditions on the system.  A natural choice is to
have a reflecting mirror surrounding the apparatus, so that we are
effectively building an acoustic black hole bomb.

\begin{figure}
\centerline{\mbox{
\psfig{figure=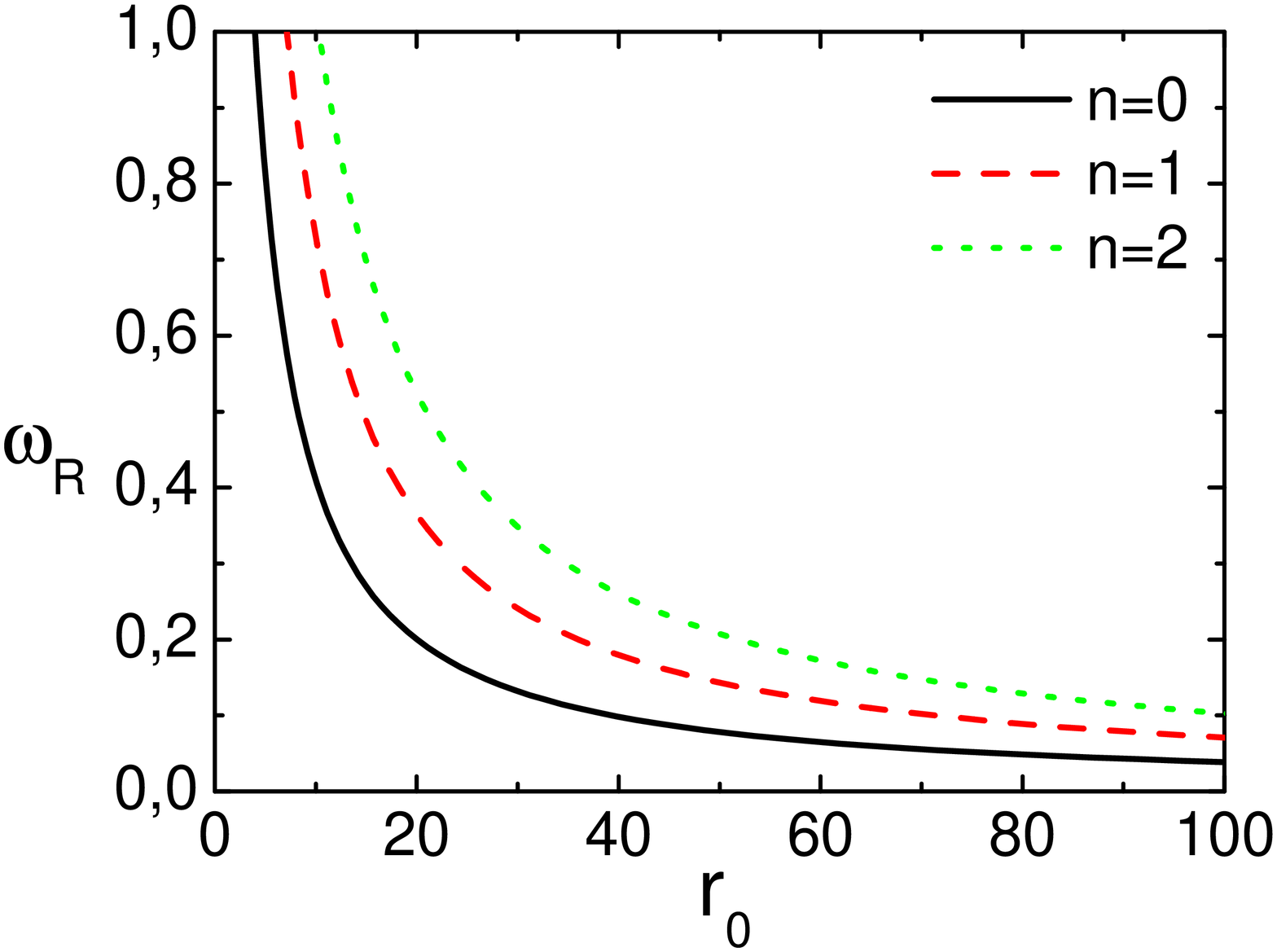,angle=270,width=9cm}
\psfig{figure=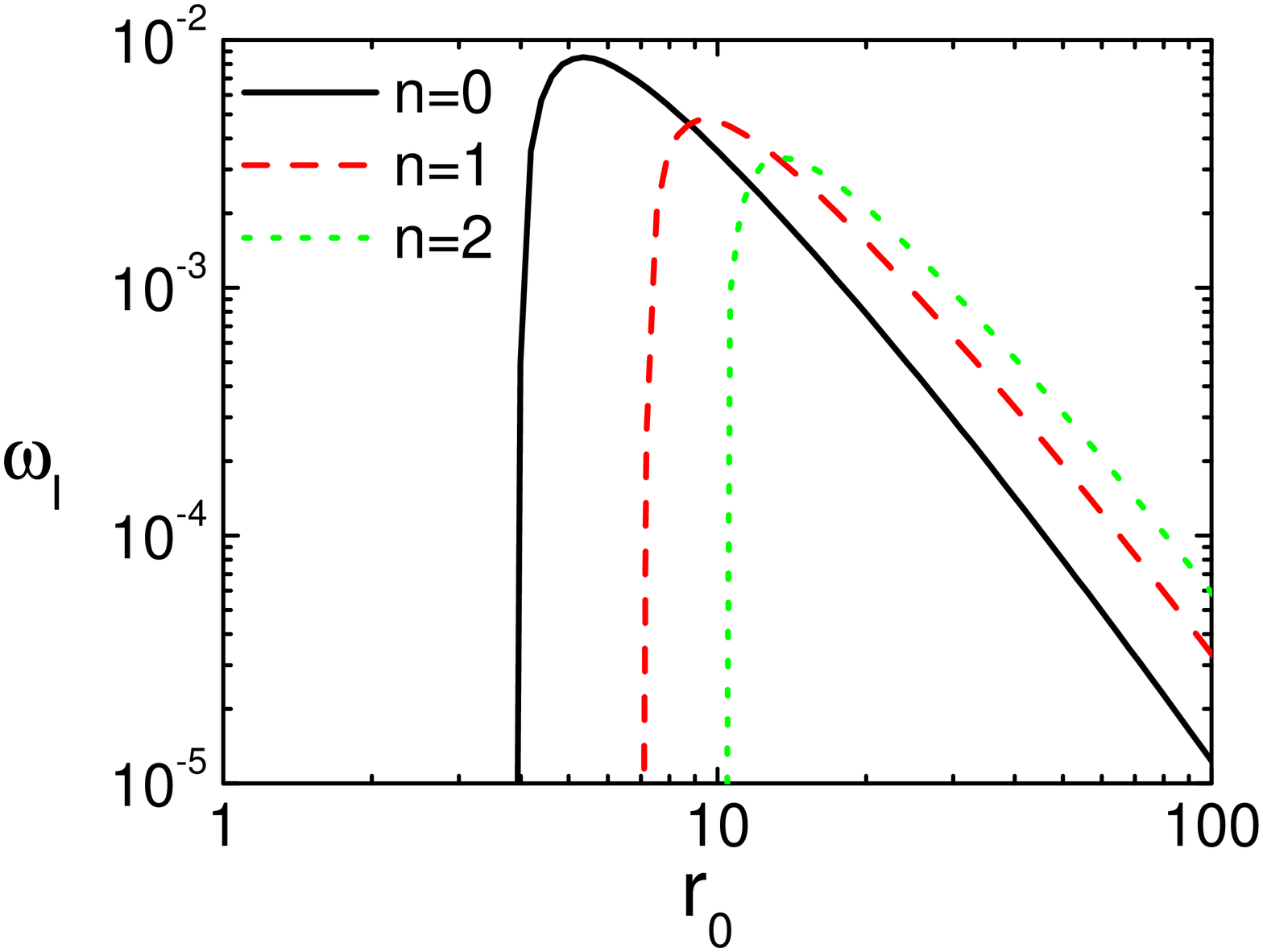,angle=270,width=9cm}}}
\caption{ Real part (left) and imaginary part (right) of the
fundamental BQN frequency and the first two overtones as a function of
mirror location $r_0$.  Both panels refer to $m=1$ and $B=1$ (but the
$B$-dependence of the real part of the BQN frequency is very weak).
With excellent accuracy $\omega_I$ crosses zero (and the instability
abruptly shuts off) at the critical radius predicted by
Eq. (\ref{bomboff}), as can be verified taking a glance at the first
row of Table \ref{tab:bessel}. }
\label{fig:BQNM1}
\end{figure}

In the following we will assume that we are in the presence of a
perfectly reflecting mirror. Correspondingly, we shall impose the
boundary condition $H=0$ at $r=r_0$. At the horizon we demand, as
usual, the presence of purely ingoing waves (waves headed towards the
horizon), i.e. $H \sim e^{-\ii(\omega-Bm)r_*}$. We are again in
presence of an eigenvalue problem. However, since the boundary
conditions have changed we shall not refer to the characteristic
frequencies as QN frequencies, but rather as Boxed QuasiNormal
frequencies (BQN frequencies). The reader is referred to \cite{bhb},
where this terminology was first introduced.  Using a direct
integration of the wave equation (see Section
\ref{integration} for details) it is quite easy to compute the BQN
frequencies. We simply integrate Eq. (\ref{waveequation2}) outwards
from the horizon, where we impose the condition (\ref{horizon}), to
the mirror location $r_0$. The (complex) BQNM frequencies are those
frequencies for which
\be
H(\omega_{BQNM},r_0)=0\,.
\ee
Some results for the fundamental BQNM and the first two overtones are
shown in Fig. \ref{fig:BQNM1}. There we fixed the black hole
rotation parameter $B=1$, and $m=1$, as a typical case. 

 From the left panel we see that the real part of the BQNM frequency
for the $n$-th overtone (where we use the convention that $n=0$
corresponds to the fundamental mode) scales as $1/r_0$, as
anticipated. The proportionality constant can be obtained by
analytical arguments similar to those in \cite{bhb}. The result is
\beq\label{BQNMr}
\omega_R=\frac{j_{m,n}}{r_0},
\eeq
where the $j_{m,n}$'s are zeros of the Bessel function of {\it
integer} order $m$ (recall that for a Kerr black hole bomb one gets
$\omega_R=j_{l+1/2,n}/r_0$ instead, cf. \cite{bhb}). The numerical
values of the $j_{m,n}$'s can be found, e.g., in \cite{abramowitz}. For
reference, we list the first few values in Table \ref{tab:bessel}.
Our numerical calculations are in excellent agreement with the
predictions of formula (\ref{BQNMr}). This is particularly true for
small values of $B$, but the $B$-dependence of the real part of the
BQN frequencies is very weak anyway, as it is for the
Kerr metric \cite{bhb}.

From the right panel we see the behavior predicted by the qualitative
arguments at the beginning of this Section: for any BQNM, the
instability is more and more efficient as the mirror radius $r_0$
becomes smaller, until eventually the mirror radius becomes small
enough that the instability shuts off.

\begin{table}
\centering
\caption{\label{tab:bessel} Zeros of the Bessel functions $j_{m,n}$
for the first few values of $m$. The real part of the BQNM frequency
is well approximated by Eq. (\ref{BQNMr}).}
\vskip 12pt
\begin{tabular}{@{}c|c|c|c@{}}  
\hline
\hline
$m$ &$j_{m,0}$ &$j_{m,1}$ &$j_{m,2}$\\
\hline
\hline
1 &3.83171 &7.01559  &10.17347\\
2 &5.13562 &8.41724  &11.61984\\
3 &6.38016 &9.76102  &13.01520\\
4 &7.58834 &11.06471 &14.37254\\
5 &8.77148 &12.33860 &15.70017\\
\hline 
\hline
\end{tabular}
\end{table}

\begin{figure}
\centerline{\mbox{
\psfig{figure=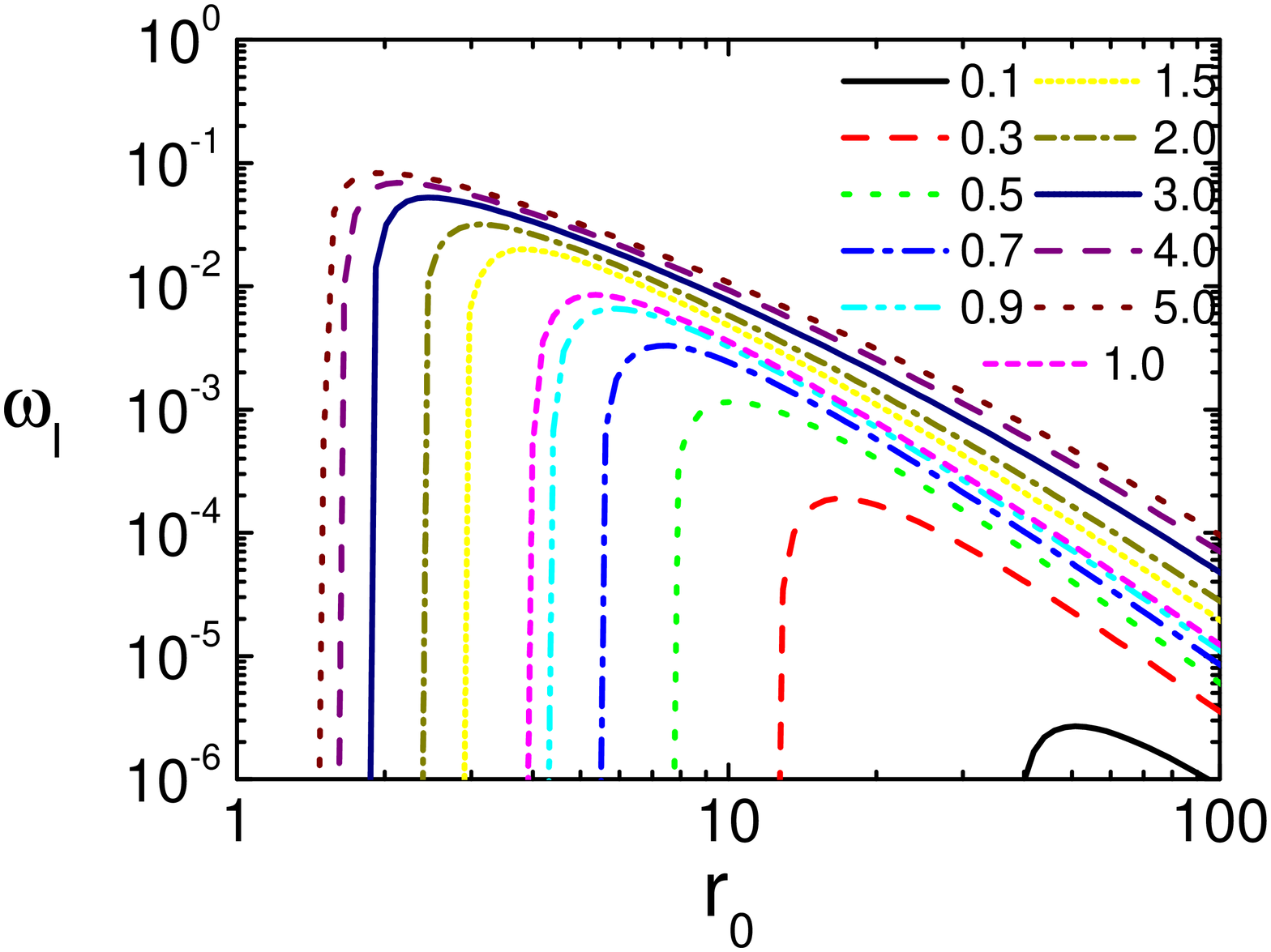,angle=270,width=9cm}
\psfig{figure=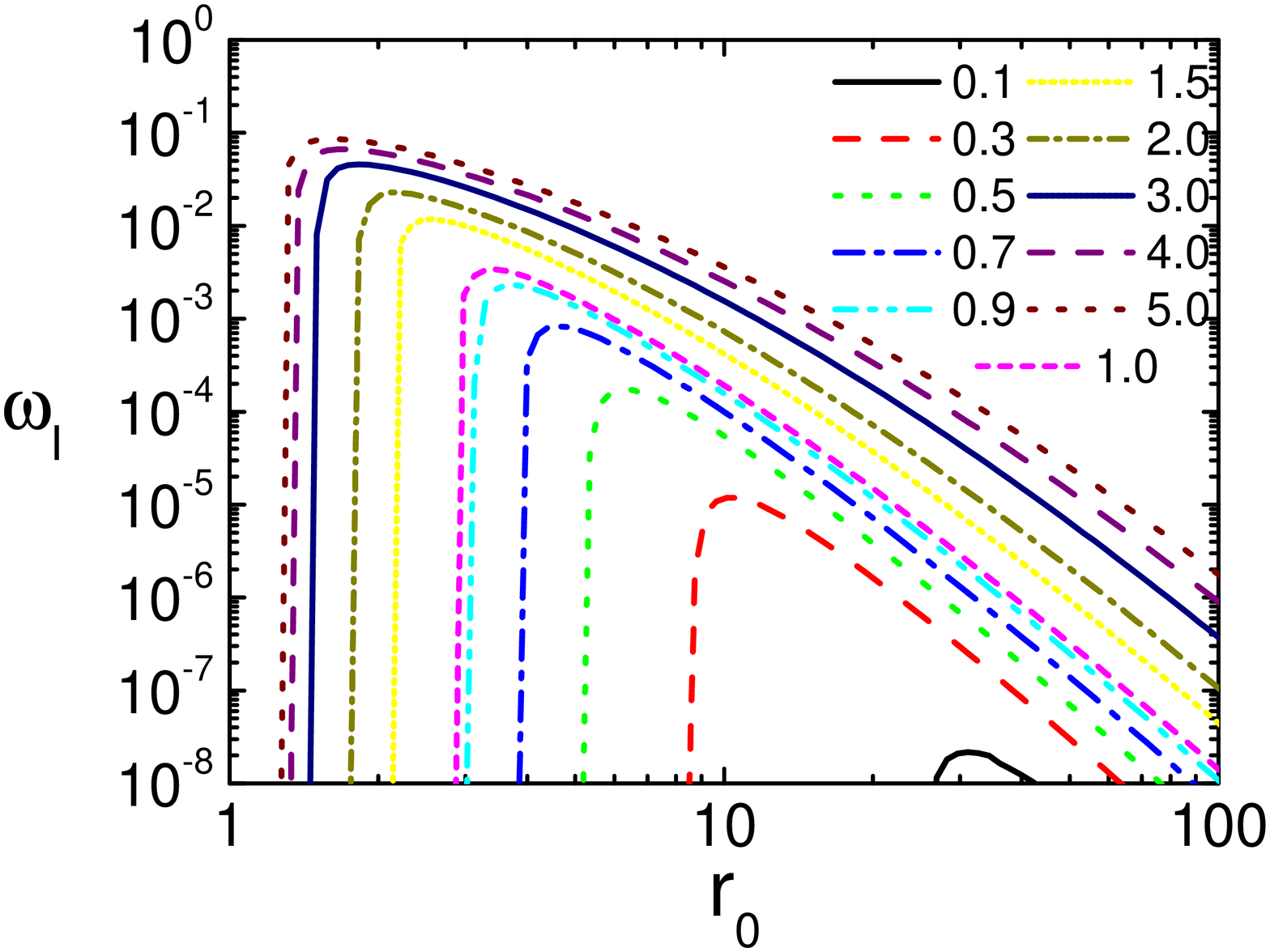,angle=270,width=9cm}}}
\caption{
Growth timescales for BQNMs, as a function of mirror location $r_0$,
for different values of $B$. The left panel refers to $m=1$, the right
panel to $m=2$. Notice the different scales for the $\omega_I$ axis.
}
\label{fig:BQNM2}
\end{figure}

A few remarks are in order: (i) The real part of the BQN frequency
(slowly) increases with overtone number $n$. (ii) Higher overtones become
stable at larger distances, but they also attain a smaller maximum
growing rate. (iii) With excellent accuracy, the instability switches
off at the critical radius predicted by the analytical formula
(\ref{BQNMr}) supplemented by the superradiance condition
(\ref{suprad}), that is:
\be
r_{0,c}\simeq \frac{j_{m,n}}{mB}\,. 
\label{bomboff}
\ee
Since $j_{m,n}$ grows linearly with $m$ (in particular, this is
asymptotically true for high $m$'s \cite{abramowitz}) the critical
radius is almost $m$-independent.  However, the instability is not so
efficient for higher $m$ as it is for small $m$, at least when
$B\lesssim 1$. This is apparent in Fig. \ref{fig:BQNM2}, where we
compare growth timescales with $m=1$ (left) and $m=2$ (right) for
acoustic black holes rotating at different rates. For example, the
maximum growth rate for $B=0.3$ is $\sim 10^{-4}$ for $m=1$, and $\sim
10^{-5}$ for $m=2$. The growth rate for $m=2$ becomes roughly
comparable to the growth rate for $m=1$ when $B>1$: look for example
at the curves corresponding to $B=5.0$. However, other physical
considerations may forbid the construction of acoustic black holes
rotating at such rates (see the arguments at the end of Section
\ref{integration}).

To get an idea of the orders of magnitude involved, let us consider
again the ``typical'' parameters for the gravity wave analogue we
introduced in Section \ref{gwanal}. Let us pick $m=1$ and an acoustic
black hole rotation parameter $\hat B=1$, which -- according to the
arguments in Section \ref{integration} -- is close to the maximum
rotation rate we may hope to achieve (recall hats distinguish
dimensionless quantities from quantities in physical units). To take
advantage of the process the mirror should be located close to the maximum
of $\omega_I$, but not quite at the maximum. For example, if we place
the mirror at $\hat r_0\simeq 10$ (cf. the right panel of
Fig. \ref{fig:BQNM1}) we get $\hat \omega_I\simeq 4\cdot 10^{-3}$ (in
this case the maximum growth rate would be $\hat \omega_I=8.5 \cdot
10^{-3}$). This corresponds to a growth time $\tau\simeq 800\,{\rm
s}\simeq 13\,{\rm minutes}$ (at the maximum we would get $\tau\simeq
6$~minutes). Roughly speaking, this means that the amplitude will
double every 13 minutes or so, and will be amplified by orders of
magnitude on timescales of the order of a few hours.  To reduce the
typical timescale to seconds, one just has to ajust the horizon radius
and wave velocity.  For example, working with $r_H \sim 0.1\,$m and
$c\sim 10$ m~s$^{-1}$, we would get a typical timescale of about 2
seconds.  This looks like a perfectly reasonable timescale to observe
an acoustic black hole bomb in the lab.

When the instability sets in the acoustic black hole loses energy and
angular momentum ($\Delta E\sim B \Delta J$). The critical radius
$r_{0,c}$ increases with decreasing $B$. To estimate the efficiency we
can use a simple argument. We can take advantage of the mechanism by
picking some large $B$ and $r_0\simeq 2 r_{0, max}$, where $r_{0,
max}$ is the radius of maximum growth timescale at the given $B$. As
the hole emits angular momentum $B$ will decrease. For any given
mirror location $r_0$, the bomb eventually switches off when condition
(\ref{bomboff}) is satisfied. From the difference in initial and final
angular momentum $\Delta B$ we can infer the extracted energy $\Delta
E$. For more details we refer to \cite{bhb}.

As a final remark we want to stress that, when we talk about
surrounding the acoustic black hole by a reflecting mirror, what we
have in mind is a generic type of mirror.  For instance, in the
shallow basin gravity wave analogue, the mirror could be a circular
rubber band of radius $r_0$ reflecting gravity waves.  A possible
alternative to implement a mirror could involve, for example,
variations in the dispersion relation for sound waves. In particular,
Sch\"utzhold and Unruh \cite{schutzhold} suggested that a simple
change of the height $h_\infty$ of the basin would imply a change of
the gravity wave speed, hence of the effective refractive index. The
changed refractive index may be used to trap the waves, which can then
be amplified via superradiant scattering until non-linear effects
dominate.

\section{The canonical non-rotating acoustic black hole}\label{canonical}

In the previous Section we considered the metric describing a rotating
$(2+1)$-dimensional acoustic black hole. Rotation implies the presence
of an ergosphere, allowing us to study in the lab some of the most
interesting phenomena concerning black hole physics (eg.
superradiance and the related rotational instabilities). However, the
non-rotating limit of the acoustic metric we considered is not the
``natural'' metric describing a non-rotating acoustic black hole. In
this Section we are going to consider another class of non-rotating,
$(3+1)$-dimensional acoustic metrics. These metrics are associated to
the most general spherically symmetric flow of an incompressible
fluid: in this sense they represent ``canonical'' metrics for a
non-rotating acoustic black hole.

\subsection{Formalism and basic equations}

It is simple to show that the acoustic metric corresponding to the
most general spherically symmetric flow of an incompressible fluid is
given by \cite{visser}
\begin{equation}
ds^2=
-c^2\left (1-\frac{r_0^4}{r^4} \right)dt^2+ 
\left (1-\frac{r_0^4}{r^4} \right)^{-1} dr^2+
r^2\left (d\theta ^2+ \sin{\theta}^2 d\phi^2 \right)\,.
\label{metriccanonical}
\end{equation}
This metric does not correspond to any of the geometries typically
considered in general relativity, but it describes (in the sense
specified above) a ``canonical'' acoustic black hole. The propagation
of small disturbances (sound waves) is again described by the massless
Klein-Gordon equation $\nabla_{\mu}\nabla^{\mu}\Psi=0$ in this
background. We can separate variables by the substitution
\be
\Psi(t,r,\phi)=\frac{\Phi(\omega,r)}{r} e^{-\ii\omega t}Y_{lm}(\theta)\,,
\ee
where $Y_{lm}(\theta)$ are the usual spherical harmonics. This yields
the wave equation
\be
\Phi_{,r_* r_*}+\left( \frac{\omega^2}{c^2}-V \right)\Phi=0\,,
\label{waveequation4}
\ee
where
\be
V =\left ( 1-\frac{r_0^4}{r^4}\right ) 
\left [ \frac{l(l+1)}{r^2}+\frac{4r_0^4}{r^6}\right ] \,,
\label{pot5}
\ee
and the tortoise coordinate $r_*$ is defined, as usual, by
$dr/dr_*=(1-r_0^4/r^4)$. Since this is a spherical symmetric problem,
the azimuthal number $m$ does not play any role, only the angular
momentum $l$ is important here.  In the next Section we summarize our
results for QNMs and wave tails of the canonical acoustic black
hole. When presenting our numerical results we will choose units such
that $c=r_0=1$. This is of course equivalent to a simple rescaling of
the radial variable ($\hat r=r/r_0$) and of the frequency ($\hat
\omega=\omega/c$), but we will omit hats in the following.

\subsection{Quasinormal modes}

A 6th order WKB analysis \cite{konoplyawkb} reveals an unstable QNM
for $l=0$. However, it is possible to prove stability of the canonical
black hole since the potential is positive-definite.  We therefore
computed the QN frequencies using first the lowest approximation
\cite{willwkb}, then 3rd order corrections \cite{will} and finally 6th
order corrections. The results are presented in Table
\ref{tab:fundnonrotcanonical}. The QN frequencies for $l=0$ and $l=1$
seem to be the problem here.  For any other $l$ the value quickly
converges as we increase the correction order of the WKB
method. Notice for example that the $l=0$ mode suffers a variation of
almost an order of magnitude as we go from the lowest approximation to
the 3rd order correction scheme, and gets unstable for the scheme
using 6th order corrections.  The reason for this failure is most
likely related to the breakdown of the basic WKB assumptions, namely
that the ratio of the derivatives of the potential to the potential
itself is small. Indeed a close inspection shows that as $l$ increases
these ratios tend to decrease. This means that the WKB method is more
reliable for higher $l$: this was first observed in the early works on
the subject \cite{willwkb,will}.  
For $l=0$ and $l=1$ the lowest WKB approximation gives the most
reliable results. This is confirmed by recent numerical work \cite{num3}.
\begin{table}
\centering
\caption{\label{tab:fundnonrotcanonical} The fundamental ($n=0$) QN
frequencies for the non-rotating canonical acoustic black hole, using
three WKB computational schemes.  $\omega _{QN}^{(1)}$ is the result
for the QN frequency using only the lowest approximation
\cite{willwkb}, $\omega _{QN}^{(3)}$ is the value obtained using 3rd
order improvements \cite{will}, and finally $\omega _{QN}^{(6)}$ was
computed using 6th order corrections \cite{konoplyawkb}.  }
\vskip 12pt
\begin{tabular}{@{}c|c|c|c@{}}  
\hline
\hline
$l$ &$\omega _{QN}^{(1)}$  &$\omega _{QN}^{(3)}$   &$\omega _{QN}^{(6)}$\\
\hline
\hline
0  & 1.13-0.73i & 0.19-1.11i &0.06+0.87i \\
1  & 1.37-0.68i & 0.53-0.71i &1.09-0.39i \\
2  & 1.82-0.64i & 1.41-0.61i &1.41-0.70i \\
3  & 2.36-0.62i & 2.10-0.62i &2.12-0.62i \\
4  & 2.94-0.62i & 2.75-0.62i &2.75-0.62i \\
\hline 
\hline
\end{tabular}
\end{table}

In the limit of large $l$ one finds
\be
\omega \sim \frac{l\sqrt{2}}{3^{3/4}}-\ii\frac{\sqrt{2}(1+2n)}{3^{3/4}}\,,
\ee
a result which agrees very well with our WKB data already for $l=3$.

The calculation of highly damped QNMs proceeds along the same lines
sketched for the $(2+1)$-dimensional acoustic black hole. In this case
the index $j=3/5$, which implies that
\be
4\pi \omega =\log{\frac{3-\sqrt{5}}{2}} -\ii(2n+1)\pi\,.
\ee
Once again asymptotic QN frequencies are not given by the logarithm of
an integer, as required in Hod's construction \cite{hod}. As we
remarked earlier this does not necessarily imply that the conjecture
is wrong, since a thermodynamical interpretation of the black hole
area is possible only when the dynamics of the system are described by
the Einstein equations.

\subsection{Late-time tails}

The analysis of late-time tails proceeds as in the case of the
$(2+1)$-dimensional acoustic black hole. We can easily show that
asymptotically the potential behaves as
\be
V \sim \frac{l(l+1)}{r_*^2}+\frac{12-5l(l+1)}{3r_*^6}\,,r\rightarrow 
\infty\,.
\ee
Following the previous analysis we now find $\alpha=6$, and since $l$
is an integer the power-law falloff is of the form
\be
\Phi \sim t^{-(2l+6)}\,.
\ee
Thus any perturbation eventually dies off as $t^{-(2l+6)}$, much more
quickly (for $m=l$) than in the $(2+1)$-dimensional acoustic black
hole background.

\section{Conclusions}\label{conclusions}

In this paper we have considered two acoustic black hole metrics: the
$(2+1)$-dimensional ``draining bathtub'' metric, and the
$(3+1)$-dimensional canonical non-rotating acoustic black hole.  We
have studied QNMs and late-time tails in both metrics, and
superradiance in the ``draining bathtub'' metric.

More specifically, we numerically computed slowly damped QNMs of the
draining bathtub metric using a third-order WKB approach for all
values of $m\neq 0$. We analytically evaluated QN frequencies in the
large-$m$ limit [Eq. (\ref{largembehav})] and set upper limits on
frequencies of unstable modes (Appendix \ref{unstable}). We showed
that highly damped modes do not tend to any simple limit. At late
times, power-law tails decay as $t^{-(2m+1)}$.  This behavior is
typical of any odd-dimensional spacetime: independently of the
presence of a black hole, if $D$ is odd the power-law falloff is
proportional to $t^{-(2l+D-2)}$ \cite{cardosoDdimtails} [the
$(2+1)$-dimensional case, $D=3$, is special: the azimuthal and angular
numbers are the same, $l=m$, since there is only one angular
coordinate]. The draining bathtub metric, possessing an ergoregion,
can superradiantly amplify waves. We computed reflection coefficients
for this superradiant scattering by numerical integration. When the
(dimensionless) acoustic black hole rotation $B=1$, the maximum
amplification is 21.2 \% for $m=1$ and 4.7 \% for $m=2$.  Enclosing
the acoustic black hole by a reflecting mirror we can destabilize the
system, making an initial perturbation grow exponentially with time:
we have an acoustic black hole bomb \cite{pressteu}, or ``dumb hole
bomb''. We computed analytically and numerically the frequencies and
growing timescales for this instability. An interesting feature of
acoustic geometries is that the acoustic black hole {\it spin} can be
varied independently of the black hole {\it mass}. Therefore, at
variance with the Kerr metric, the spin can be made (at least in
principle) arbitrarily large, and rotational superradiance in acoustic
black holes can be very efficient. Gravity waves in shallow water
\cite{schutzhold} provide a concrete example of an experimental setup
for studying classical physics in a draining bathtub metric [and of
experimental difficulties in increasing arbitrarily the rotation rate:
see the discussion leading to Eq. (\ref{Blimit})]. In this case, and
for a typical choice of parameters, growth times for the black hole
bomb instability would be of the order of minutes: this appears to be
well within the range of experimental possibilities. Finally one can
speculate, following \cite{basak2}, that superfluid HeII could be used
to observe some quantum effects, such as a discrete superradiant
amplification.

For the canonical acoustic black hole we computed slowly-damped QNMs
using first, third and sixth order WKB. For $l<2$ the WKB method does
not seem to converge, and even yields an unstable frequency at 6th
order when $l=0$. This is only due to bad convergence properties of
the WKB technique, since the canonical acoustic black hole is
stable. In the high-damping limit QNMs are given by $4\pi\omega =
\log{[(3-\sqrt{5})/2]}-\ii(2n+1)\pi$, so they are not proportional to
the logarithm of an integer, as required by recent conjectures. 
This is not surprising since there are no Einstein equations
for acoustic metrics, i.e., the acoustic metric evolution
is not governed by Einstein's equations.
Therefore, acoustic black holes can teach us a lot about quantum gravity
\cite{visser,corleyjacobson}, but cannot shed any light on its possible
connection with highly damped quasinormal modes.

The late time falloff in the canonical acoustic black hole metric 
is proportional to $t^{-(2l+6)}$,
to be compared with the $t^{-(2l+3)}$ decay of 4-dimensional
Schwarzschild black holes and with the $t^{-(2l+3D-8)}$ decay of
even-dimensional spherically symmetric black holes with $D>4$
\cite{cardosoDdimtails}.

\section*{Acknowledgements}

We are grateful to Carlos Barcel\'o and Ted Jacobson for useful
discussions.  This work was partially funded by Funda\c c\~ao para a
Ci\^encia e Tecnologia (FCT) -- Portugal through project
CERN/FNU/43797/2001.  V.C. acknowledges financial support from FCT
through grant SFRH/BPD/2003.
This work was supported in part by the National Science Foundation under
grant PHY 03-53180.

\appendix

\section{Upper bounds for quasinormal frequencies of unstable modes}
\label{unstable}

To derive some bounds on the magnitude of QN frequencies of possible
unstable QNMs we shall follow Detweiler and Ipser \cite{det}.  Let us
begin with the Klein-Gordon equation for the evolution of the field,
$\nabla_{\mu}\nabla^{\mu}\Psi=0$.  Using the metric (\ref{metric2})
and performing a mode decomposition $\Psi(t,r,\phi)=R(r)e^{\ii
(m\phi-\omega t)}$, this can be written as
\be
\frac{r}{fc^2}\omega ^2 R+\frac{d}{dr}\left[ rfR'\right]-
\frac{2Bm\omega}{c^2fr}R-
\left (\frac{1}{r}-\frac{B^2}{c^2r^3f}\right )m^2R=0\,,
\label{eq1}
\ee
where we have defined $f\equiv \Delta^{-1}=1-A^2/c^2r^2$.

Multiply by $R^*$ and integrate from the horizon to spatial
infinity. The result is
\be
\int_{r_H} ^{\infty}dr \left[ \frac{r}{fc^2}\omega ^2 |R| ^2-rf|R'| ^2-
\frac{2Bm\omega}{c^2fr}|R| ^2-
\left (\frac{1}{r}-\frac{B^2}{c^2r^3f}\right )m^2|R|^2 \right ]=0\,,
\label{eq2}
\ee
where we have used an integration by parts and discarded the surface
integrals.  In fact, for unstable modes (and for these only) the
boundary conditions guarantee that an unstable mode vanishes
exponentially as $r \rightarrow r_H$ or $r\rightarrow \infty$.  The
imaginary part of this equation yields
\be
\int_{r_H} ^{\infty}dr \left( r^2 \omega_R -Bm
 \right )\frac{|R| ^2}{frc^2}=0\,, 
\label{eq3}
\ee
where we used $\omega=\omega_R+{\rm i}\,\omega_I$. 
Therefore $\omega_R$ and $Bm$ must have the same sign for unstable
modes.  Furthermore, since $|R| ^2/frc^2$ is always positive, the
quantity $(r^2 \omega_R -Bm)$ must be negative somewhere for the
integral to vanish. Since $r^2 \omega_R$ increases with $r$, if this
quantity is somewhere negative, then it certainly is negative at the
horizon. In this way we get an upper bound for $\omega_R$:
\be
\omega_R<\frac{Bm}{r_H^2}\,.
\label{eq4}
\ee
To get a similar bound on $\omega_I$ consider now the real part of
Eq. (\ref{eq2}):
\be
\int_{r_H} ^{\infty}dr \left[ \omega_I ^2-\omega_R^2+\frac{2Bm\omega_R}{r^2}
+\frac{c^2fm^2}{r^2}-\frac{B^2m^2}{r^4}+
f^2c^2\frac{|R'| ^2}{|R| ^2}
 \right ]\frac{r |R| ^2}{fc^2}=0\,.
\label{eq5}
\ee
The positive part of the integrand is 
(assuming without loss of generality that $Bm\omega_R>0$)
\be
\omega_I ^2+\frac{2Bm\omega_R}{r^2}
+\frac{c^2fm^2}{r^2}+
f^2c^2\frac{|R'| ^2}{|R| ^2}>\omega_I ^2\,.
\label{eq6}
\ee
The negative part is maximized at the horizon:
\be
\omega_R^2+\frac{B^2m^2}{r^4}<\omega_R^2+\frac{B^2m^2}{r_H^4}\,.
\label{eq7}
\ee
A necessary condition for the integral to vanish is that the integrand
be negative at the horizon, and this implies
\be
\omega_I^2<\omega_R^2+\frac{B^2m^2}{r_H^4}\,,
\label{eq8}
\ee
Using the bound (\ref{eq4}) for $\omega_R$ we finally get an upper
bound for $\omega_I$:
\be
\omega_I^2<\frac{2B^2m^2}{r_H^4}\,.
\label{eq9}
\ee
Notice that the upper bound for $\omega_R$ ensures we are in the
superradiant (or better, superresonant) regime. This seems to be a
general feature, since it has been shown to hold in several black hole
spacetimes. If the instability sets in at all, the system behaves as a
black hole bomb.



\begin{thebibliography}{99}

\bibitem{hawking} S. W. Hawking, Nature {\bf 248}, 30 (1974);
S. W. Hawking, Commun. Math. Phys. {\bf 43}, 199 (1975).

\bibitem{visserlaws} M. Visser,
Phys. Rev. Lett. {\bf 80}, 3436 (1998).

\bibitem{visserhawking} M. Visser,
Int. J. Mod. Phys. D {\bf 12}, 649 (2003).

\bibitem{giddings} S. B. Giddings, S. Thomas, Phys. Rev. D {\bf 65},
056010 (2002).

\bibitem{unruh} W. G. Unruh, Phys. Rev. Lett. {\bf 46}, 1351 (1981).

\bibitem{novello} M. Novello, M. Visser and G. Volovik (editors), {\it
Artificial black holes} (World Scientific, Singapore, 2002).

\bibitem{visser} M. Visser, Class. Quantum Grav. {\bf 15}, 1767 (1998).

\bibitem{barcelo}  C. Barcel\'o, S. Liberati and M. Visser,
Int. J. Mod. Phys. {\bf A} 18, 3735 (2003).

\bibitem{unruh2} R. Schutzhold and W. G. Unruh, quant-ph/0408145.

\bibitem{corleyjacobson} S. Corley and T. Jacobson, Phys. Rev. D {\bf
59}, 124011 (1999).

\bibitem{schwinger} 
S. Detweiler, Phys. Rev. D {\bf 22}, 2323 (1980); T. J. Zouros and
D. M. Eardley, Annals Phys. {\bf 118}, 139 (1979); T. Damour,
N. Deruelle and R. Ruffini, Lett. Nuovo Cim. {\bf 15}, 257 (1976).

\bibitem{hod} S. Hod, 
Phys. Rev. Lett. {\bf 81}, 4293 (1998).

\bibitem{kokkotas} K. D. Kokkotas and B. G. Schmidt, 
Living Rev. Rel. {\bf2}, 2 (1999);
H.-P. Nollert, 
Class. Quantum Grav. {\bf 16}, R159 (1999).

\bibitem{echeverria} F. Echeverria,
Phys. Rev. D {\bf 40}, 3194 (1989);
L. S. Finn, 
Phys. Rev. D {\bf 46}, 5236 (1992).

\bibitem{price1}
R. H. Price, 
Phys. Rev. D{\bf 5}, 2419 (1972).

\bibitem{schutzhold}
R. Sch\"utzhold, W. G. Unruh,
Phys. Rev. D {\bf 66}, 044019 (2002).

\bibitem{basak1} S. Basak and P. Majumdar, 
Class. Quant. Grav. {\bf 20}, 2929 (2003). 

\bibitem{basak2} S. Basak and P. Majumdar, 
Class. Quant. Grav. {\bf 20}, 3907 (2003). 

\bibitem{pressteu} W. Press and S. Teukolsky,
Astrophys. J. {\bf 193}, 443 (1974).

\bibitem{bhb} V. Cardoso, O. J. C. Dias, J. P.S. Lemos and S. Yoshida, 
Phys. Rev. D{\bf 70}, 044039 (2004); {\bf 70}, 044039(E)(2004). 

\bibitem{putten} For a possible astrophysical relevance of the
black hole bomb mechanism, see
M. H. P. M. Putten,
Science {\bf 284}, 115 (1999);
A. Aguirre, 
Astrophys. J. {\bf 529}, L9 (2000).

\bibitem{PGL} P. Painlev\'e, C. R. Hebd. Seances Acad. Sci. {\bf 173},
677 (1921); A. Gullstrand, Ark. Mat. Astron. Fys. {\bf 16}, 1 (1922);
G. Lema\^itre, Ann. Soc. Sci. Bruxelles, Ser. 1 {\bf 53}, 51 (1933).

\bibitem{chandra} S. Chandrasekhar, {\it Hydrodynamic and
Hydromagnetic Stability} (Dover Publications, New York, 1981).

\bibitem{drazin} P. G. Drazin and W. H. Reid, {\it Hydrodynamic
Stability} (Cambridge University Press, 2004).

\bibitem{willwkb} B. F. Schutz and C. M. Will,
Astrophys. Journal {\bf 291}, L33 (1985);

\bibitem{will} C. M. Will and S. Iyer,
Phys. Rev. D {\bf 35}, 3621 (1987);
S. Iyer,
Phys. Rev. D {\bf 35}, 3632 (1987);
E. Seidel and Sai Iyer, 
Phys.Rev. D {\bf 41}, 374 (1990).

\bibitem{kokkotaswkb} K. D. Kokkotas, 
Class. Quant. Grav. {\bf 8}, 2217 (1991).

\bibitem{konoplyawkb} R. A. Konoplya,
Phys. Rev. D {\bf 68}, 024018 (2003).

\bibitem{ferrari} V. Ferrari and B. Mashhoon,
Phys. Rev. {\bf D30}, 295 (1984).

\bibitem{dreyerfinn} O. Dreyer, B. Kelly, B. Krishnan, L. S. Finn,
D. Garrison and R. Lopez-Aleman, Class. Quantum Grav. {\bf 21}, 787
(2004).

\bibitem{cardoso4} V. Cardoso, J. P. S. Lemos and S. Yoshida,
gr-qc/0410107.

\bibitem{dreyer} O. Dreyer,
Phys. Rev. Lett. {\bf 90}, 081301 (2003). 

\bibitem{motl1} L. Motl,
Adv. Theor. Math. Phys. {\bf 6}, 1135 (2003).

\bibitem{motl2} L. Motl and A. Neitzke,
Adv. Theor. Math. Phys. {\bf 7}, 2 (2003).

\bibitem{num1} H.-P. Nollert, 
Phys. Rev. D {\bf 47}, 5253 (1993);
E. Berti and K. D. Kokkotas,  
Phys. Rev. D {\bf 68}, 044027 (2003).

\bibitem{num3} V. Cardoso, J. P. S. Lemos and S. Yoshida
Phys. Rev. D {\bf 69}, 044004 (2004). 

\bibitem{nils} N. Andersson and C. J. Howls,
Class. Quantum Grav. {\bf 21}, 1623 (2004).

\bibitem{ricardo} V. Cardoso, J. Nat\'ario and R. Schiappa,
J. Math. Phys. (in press); hep-th/0403132.

\bibitem{tamaki} T. Tamaki and H. Nomura,
Phys. Rev. D {\bf 70}, 044041 (2004);
J. Kettner, G. Kunstatter and A.J.M. Medved;
gr-qc/0408042;
S. Chen and J. Jing, gr-qc/0409013. 

\bibitem{num2} V. Cardoso, R. Konoplya and J. P. S. Lemos,
Phys. Rev. D {\bf 68}, 044024 (2003);
S. Yoshida and T. Futamase,
Phys. Rev. D {\bf 69}, 064025 (2004).
R. A. Konoplya and A. Zhidenko,
JHEP {\bf 0406}, 037 (2004).

\bibitem{neitzke} We thank Andrew Neitzke for sharing with us a
related unpublished calculation for the Kerr metric.

\bibitem{TedEinsteinEOS} T. Jacobson, Phys. Rev. Lett. {\bf 75}, 1260
(1995).

\bibitem{BLSV} C. Barcel\'o, S. Liberati, S. Sonego and M. Visser,
gr-qc/0408022.

\bibitem{price2}
C. Gundlach, R. H. Price and J. Pullin, 
Phys. Rev. D{\bf 49}, 883 (1994);
C. Gundlach, R. H. Price and J. Pullin, 
Phys. Rev. D{\bf 49}, 890 (1994);
S. Hod,
Phys. Rev. D{\bf 58}, 104022 (1998);
L. Barack and A. Ori,
Phys. Rev. Lett. {\bf 82}, 4388 (1999);
N. Andersson and K. Glampedakis,
Phys. Rev. Lett. {\bf 84}, 4537 (2000);
H. Koyama and A. Tomimatsu,
Phys. Rev. D{\bf 64}, 044014 (2001);
S. Hod,
Class. Quant. Grav. {\bf 18}, 1311 (2001).

\bibitem{ching1} E. S. C. Ching, P. T. Leung, W. M. Suen and K. Young,
Phys. Rev. Lett. {\bf 74}, 2414 (1995). 

\bibitem{ching2} E. S. C. Ching, P. T. Leung, W. M. Suen and K. Young,
Phys. Rev. D{\bf 52}, 2118 (1995). 

\bibitem{cardosoDdimtails} V. Cardoso, S. Yoshida, O. J. C. Dias and
J. P.S. Lemos, Phys. Rev. D {\bf 68}, 061503 (2003).

\bibitem{ginzburg} V. L. Ginzburg and I. M. Frank, Dokl. Akad. Nauk
SSSR {\bf 56}, 583 (1947); for a recent review cf. V. L. Ginzburg, in
{\it Progress in Optics XXXII}, edited by E. Wolf (Elsevier,
Amsterdam, 1993).

\bibitem{bekschiffer}
J. D. Bekenstein and M. Schiffer,
Phys. Rev. D {\bf 58}, 064014 (1998).

\bibitem{zeldovich} 
Ya. B. Zel'dovich, JETP Lett. {\bf 14}, 180 (1971); Sov. Phys. JETP
{\bf 35}, 1085 (1972).

\bibitem{superr}
C. W. Misner,
Phys. Rev. Lett. {\bf 28}, 994 (1972);
A. A. Starobinsky,
Sov. Phys. JETP {\bf 37}, 28 (1973);
A. A. Starobinsky and S. M. Churilov,
Sov. Phys. JETP {\bf 38}, 1 (1973);
W. Unruh,
Phys. Rev. D {\bf 10}, 3194 (1974);
W. H. Press and S. A. Teukolsky,
Astrophys. Journal {\bf 185}, 649 (1973).

\bibitem{beksuperr}
J. D. Bekenstein,
Phys. Rev. D {\bf 7}, 949 (1973).

\bibitem{penrose} 
R. Penrose, Nuovo Cimento {\bf 1}, 252 (1969).

\bibitem{friedman}
J. L. Friedman,
Commun. Math. Phys. {\bf 63}, 243 (1978).

\bibitem{cominsschutz}
N. Comins and B. F. Schutz,
Proc. R. Soc. Lond. A {\bf 364}, 211 (1978).

\bibitem{ptbhb}
W. H. Press and S. A. Teukolsky,
Nature {\bf 238}, 211 (1972).

\bibitem{tedvolovik}
T. Jacobson and G. E. Volovik,
Phys. Rev. D {\bf 58}, 064021 (1998).

\bibitem{alp} 
N. Andersson, P. Laguna and P. Papadopoulos, Phys. Rev. D {\bf 58},
087503 (1998).

\bibitem{abramowitz} 
M. Abramowitz and A. Stegun, {\it Handbook of mathematical functions}
(Dover Publications, New York, 1970).

\bibitem{det} S. L. Detweiler and J. R. Ipser,
Astrophys. Journal {\bf 185}, 675 (1973).
\end{thebibliography}
\end{document}